\documentclass[iop]{emulateapj}
\usepackage{subfigure}
\usepackage{multirow}
\usepackage{gensymb}

\makeatletter
\newcommand\ionpat[2]{#1$\;${\scshape{#2}}}

\def\one{111$^{\circ}$}
\def\two{119$^{\circ}$}

\def\halphaabsorbminimumvelgrism{$-8356$}
\def\halphaabsorbminimumvelstdgrism{1105}
\def\halphaabsorbminimumwavegrism{15,880}
\def\halphaabsorbminimumwavestdgrism{60}

\def\chisqPCygnigrism{167.4}
\def\numparPCygnigrism{6}
\def\chisqGaussiangrism{181.0}
\def\numparGaussiangrism{3}
\def\chisqLorentziangrism{185.8}
\def\numparLorentziangrism{3}
\def\AICGaussiangrism{$-4.8$}
\def\AICLorentziangrism{0.0}
\def\AICPCygnigrism{$-12.4$}

\def\chisqGaussianvlt{7206.5}
\def\numparGaussianvlt{5}
\def\chisqLorentzianvlt{7223.8}
\def\numparLorentzianvlt{5}
\def\AICGaussianvlt{$-17.3$}
\def\AICLorentzianvlt{0.0}

\def\pvalueabsorptiongrism{0.004}

\def\halphaabsorbminimumwavevlt{15,983}
\def\halphaabsorbminimumwavestdvlt{159}
\def\halphaabsorbminimumvelvlt{$-6465$}
\def\halphaabsorbminimumvelstdvlt{2918}

\def\pvalueabsorptionvlt{0.800}

\def\starmag{$r=15.1$ mag AB}

\def\binwidth{100\,\AA}

\def\grismphot{December 28--31, 2014}

\def\grismstart{December 23, 2014}

\def\grismmid{December 29, 2014}
 
\def\grismend{January 5, 2015}

\def\grism{December 23, 2014 through January 5, 2015}

\def\mosfirephase{$-62\pm8$\,d}
\def\grismphase{$-47\pm8$\,d}
\def\xshooterphase{$+16\pm8$\,d}

\def\ejectamass{$20\pm5$~M$_{\odot}$}

\def\magonesixty{25.1 mag AB}

\newcommand{\hb}{H$\beta$} 
\newcommand{\ha}{H$\alpha$} 
\newcommand{\hg}{H$\gamma$}

\newcommand{\sii}{[\ion{S}{2}]} 
\newcommand{\oii}{[\ion{O}{2}]} 
\newcommand{\oiii}{[\ion{O}{3}]}
\newcommand{\nii}{[\ion{N}{2}]} 
\newcommand{\mgii}{\ion{Mg}{2}} 
\newcommand{\baii}{\ion{Ba}{2}} 
\newcommand{\farc}{\hbox{$.\!\!^{\prime\prime}$}} 

\submitted{Submitted to The Astrophysical Journal}

\begin{document}
\title{SN Refsdal: Classification as a Luminous and Blue SN~1987A-like Type II Supernova}
\shorttitle{Spectroscopic Classification of SN Refsdal}

\newcommand{\HubbleFellow}{Hubble Fellow}
\newcommand{\Packard}{Packard Fellow}
\newcommand{\CalTech}{California Institute of Technology, 1200 East California Boulevard, Pasadena, CA 91125}
\newcommand{\Cantabria}{IFCA, Instituto de F\'isica de Cantabria (UC-CSIC), Av. de Los Castros s/n, 39005 Santander, Spain}
\newcommand{\IFCA}{\Cantabria}
\newcommand{\JHU}{Department of Physics and Astronomy, The Johns Hopkins University, 3400 N. Charles St., Baltimore, MD 21218, USA}
\newcommand{\Michigan}{Department of Astronomy, University of Michigan, 1085 S. University Avenue, Ann Arbor, MI 48109, USA}
\newcommand{\UCDavis}{University of California, Davis, 1 Shields Avenue, Davis, CA 95616}
\newcommand{\UCLA}{University of California, Los Angeles, CA 90095}
\newcommand{\USC}{Department of Physics and Astronomy, University of South Carolina, 712 Main St., Columbia, SC 29208, USA}
\newcommand{\RCEU}{Research Center for the Early Universe, University of Tokyo, 7-3-1 Hongo, Bunkyo-ku, Tokyo 113-0033, Japan}
\newcommand{\TokyoPhys}{Department of Physics, University of Tokyo, 7-3-1 Hongo, Bunkyo-ku, Tokyo 113-0033, Japan}
\newcommand{\IPMU}{Kavli Institute for the Physics and Mathematics of the Universe (Kavli IPMU, WPI), University of Tokyo, 5-1-5 Kashiwanoha, Kashiwa, Chiba 277-8583, Japan}
\newcommand{\TokyoAstro}{Department of Astronomy, Graduate School of Science, The University of Tokyo, 7-3-1 Hongo, Bunkyo-ku, Tokyo 113-0033, Japan}
\newcommand{\DARK}{Dark Cosmology Centre, Niels Bohr Institute, University of Copenhagen, Juliane Maries Vej 30, DK-2100 Copenhagen, Denmark} 
\newcommand{\INFN}{INFN, Sezione di Bologna, Viale Berti Pichat 6/2, I-40127 Bologna, Italy}
\newcommand{\EHU}{Fisika Teorikoa, Zientzia eta Teknologia Fakultatea, Euskal Herriko Unibertsitatea UPV/EHU}
\newcommand{\Basque}{IKERBASQUE, Basque Foundation for Science, Alameda Urquijo, 36-5 48008 Bilbao, Spain}
\newcommand{\Berkeley}{Department of Astronomy, University of California, Berkeley, CA 94720-3411, USA}
\newcommand{\STScI}{Space Telescope Science Institute, 3700 San Martin Dr., Baltimore, MD 21218, USA}
\newcommand{\Ferrara}{Dipartimento di Fisica e Scienze della Terra, Universit\`{a} degli Studi di Ferrara, via Saragat 1, I-44122, Ferrara, Italy}
\newcommand{\INAF}{INAF, Osservatorio Astronomico di Bologna, via Ranzani 1, I-40127 Bologna, Italy}
\newcommand{\UCSB}{Department of Physics, University of California, Santa Barbara, CA 93106-9530, USA}
\newcommand{\SantaBarbara}{\UCSB}
\newcommand{\Kapteyn}{Kapteyn Astronomical Institute, University of Groningen, Postbus 800, 9700 AV Groningen, the Netherlands}
\newcommand{\WKU}{Department of Physics, Western Kentucky University, Bowling Green, KY 42101, USA}
\newcommand{\IAP}{Institut d’Astrophysique de Paris, UMR7095 CNRS-Universit\'{e} Pierre et Marie Curie, 98bis bd Arago, F-75014 Paris, France}
\newcommand{\ASIAA}{Institute of Astronomy and Astrophysics, Academia Sinica, P.O. Box 23-141, Taipei 10617, Taiwan}
\newcommand{\TokyoKashiwa}{Institute for Cosmic Ray Research, The University of Tokyo, Kashiwa, Chiba 277-8582, Japan}
\newcommand{\Munich}{University Observatory Munich, Scheinerstrasse 1, D-81679 Munich, Germany} 
\newcommand{\KICPStanford}{Kavli Institute for Particle Astrophysics and Cosmology, Stanford University, 452 Lomita Mall, Stanford, CA 94305, USA}
\newcommand{\Andalucia}{Instituto de Astrof\'isica de Andaluc\'ia (CSIC), E-18080 Granada, Spain}
\newcommand{\SaoPaulo}{Instituto de Astronomia, Geof\'isica e Ci\^encias Atmosf\'ericas, Universidade de S\~ao Paulo, Cidade Universit\'aria, 05508-090, S\~ao Paulo, Brazil}
\newcommand{\AMNH}{Department of Astrophysics, American Museum of Natural History, Central Park West and 79th Street, New York, NY 10024, USA}
\newcommand{\NYU}{Center for Cosmology and Particle Physics, New York University, New York, NY 10003, USA}
\newcommand{\Arizona}{Department of Astronomy, University of Arizona, Tucson, AZ 85721, USA}
\newcommand{\Rutgers}{Department of Physics and Astronomy, Rutgers, The State University of New Jersey, Piscataway, NJ 08854, USA}
\newcommand{\NOAO}{National Optical Astronomical Observatory, Tucson, AZ 85719, USA}
\newcommand{\LCOGT}{Las Cumbres Observatory Global Telescope Network, 6740 Cortona Dr., Suite 102, Goleta, California 93117, USA}
\newcommand{\IllinoisAstro}{ Astronomy Department, University of Illinois at Urbana-Champaign, 1002 W.\ Green Street, Urbana, IL 61801, USA }
\newcommand{\IllinoisPhysics}{ Department of Physics, University of Illinois at Urbana-Champaign, 1110 W.\ Green Street, Urbana, IL 61801, USA }
\newcommand{\UTAustin}{Department of Astronomy, University of Texas at Austin, Austin, TX 78712, USA}

\newcounter{affilct}
\setcounter{affilct}{0}

\makeatletter
\newcommand{\affilref}[1]{%
  \@ifundefined{c@#1}%
    {\newcounter{#1}%
     \setcounter{#1}{\theaffilct}%
     \refstepcounter{affilct}%
     \label{#1}%
     }{}%
  \ref{#1}%
 }
\makeatother

\makeatletter
\newcommand*\affilreftxt[2]{%
  \@ifundefined{c@#1txt}
    {\newcounter{#1txt}%
     \setcounter{#1txt}{1}
     \altaffiltext{\ref{#1}}{#2}
     }{
     }
  }
\makeatother

\author{P.~L.~Kelly\altaffilmark{\affilref{Berkeley}}}
\affilreftxt{Berkeley}{\Berkeley}
\email{pkelly@astro.berkeley.edu}

\author{G.~Brammer\altaffilmark{\affilref{STScI}}}
\affilreftxt{STScI}{\STScI}

\author{J.~Selsing\altaffilmark{\affilref{DARK}}}
\affilreftxt{DARK}{\DARK}

\author{R.~J.~Foley\altaffilmark{\affilref{IllinoisPhysics},\affilref{IllinoisAstro}}}
\affilreftxt{IllinoisPhysics}{\IllinoisPhysics}
\affilreftxt{IllinoisAstro}{\IllinoisAstro}

\author{J.~Hjorth\altaffilmark{\affilref{DARK}}}
\affilreftxt{DARK}{\DARK}

\author{S.~A.~Rodney\altaffilmark{\affilref{USC},}}
\affilreftxt{USC}{\USC}

\author{L.~Christensen\altaffilmark{\affilref{DARK}}}
\affilreftxt{DARK}{\DARK}

\author{L.-G.~Strolger\altaffilmark{\affilref{STScI}}}
\affilreftxt{STScI}{\STScI}

\author{A.~V.~Filippenko\altaffilmark{\affilref{Berkeley}}}
\affilreftxt{Berkeley}{\Berkeley}

\author{T.~Treu\altaffilmark{\affilref{UCLA},\affilref{Packard}}}
\affilreftxt{UCLA}{\UCLA}
\affilreftxt{Packard}{\Packard}

\author{C.~C.~Steidel\altaffilmark{\affilref{CalTech}}}
\affilreftxt{CalTech}{\CalTech}

\author{A.~Strom\altaffilmark{\affilref{CalTech}}}
\affilreftxt{CalTech}{\CalTech}

\author{A.~G.~Riess\altaffilmark{\affilref{JHU},\affilref{STScI}}}
\affilreftxt{JHU}{\JHU}
\affilreftxt{STScI}{\STScI}

\author{A.~Zitrin\altaffilmark{\affilref{CalTech},\affilref{HubbleFellow}}}
\affilreftxt{CalTech}{\CalTech}

\author{K.~B.~Schmidt\altaffilmark{\affilref{UCSB}}}
\affilreftxt{UCSB}{\UCSB}

\author{M.~Brada\v{c}\altaffilmark{\affilref{UCDavis}}}
\affilreftxt{UCDavis}{\UCDavis}

\author{S.~W.~Jha\altaffilmark{\affilref{Rutgers}}}
\affilreftxt{Rutgers}{\Rutgers}

\author{M.~L.~Graham\altaffilmark{\affilref{Berkeley}}}
\affilreftxt{Berkeley}{\Berkeley}

\author{C.~McCully\altaffilmark{\affilref{LCOGT},\affilref{UCSB}}}
\affilreftxt{LCOGT}{\LCOGT}
\affilreftxt{UCSB}{\UCSB}

\author{O.~Graur\altaffilmark{\affilref{NYU},\affilref{AMNH}}}
\affilreftxt{NYU}{\NYU}
\affilreftxt{AMNH}{\AMNH}

\author{B.~J.~Weiner\altaffilmark{\affilref{Arizona}}}
\affilreftxt{Arizona}{\Arizona}

\author{J.~M.~Silverman\altaffilmark{\affilref{UTAustin}}}
\affilreftxt{UTAustin}{\UTAustin}


\begin{abstract}
We have acquired {\it Hubble Space Telescope} {\it (HST)} and Very Large Telescope near-infrared spectra and images  of supernova (SN) Refsdal after its discovery as an Einstein cross in Fall 2014. 
The {\it HST} light curve of SN Refsdal matches the distinctive, slowly rising light curves of SN~1987A-like supernovae (SNe), and we find strong evidence for a broad H$\alpha$ P-Cygni profile in the {\it HST} grism spectrum at the redshift ($z=1.49$) of the spiral host galaxy. SNe~IIn, powered by circumstellar interaction, could provide a good match to the light curve of SN Refsdal, but the spectrum of a SN IIn would not show broad and strong H$\alpha$ absorption.  From the grism spectrum, we measure an H$\alpha$ expansion velocity consistent with those of SN~1987A-like SNe at a similar phase. The luminosity, evolution, and Gaussian profile of the H$\alpha$ emission of the WFC3 and X-shooter spectra, separated by $\sim 2.5$ months in the rest frame, provide additional evidence that supports the SN~1987A-like classification.
In comparison with other examples of SN~1987A-like SNe, SN Refsdal has a blue $B-V$ color and a high luminosity for the assumed range of potential magnifications.
If SN Refsdal can be modeled as a scaled version of SN~1987A, we estimate it would have an ejecta mass of \ejectamass.
The evolution of the light curve at late times will provide additional evidence about the potential existence of any substantial circumstellar material (CSM). 
Using MOSFIRE and X-shooter spectra, we estimate a subsolar host-galaxy metallicity ($8.3 \pm 0.1$ dex and $<$8.4 dex, respectively) near the explosion site. 
\end{abstract}

\maketitle 

\section{Introduction}
\citet{refsdal64} first considered the possibility that a gravitational lens might create 
multiple images of a background supernova (SN) explosion. He showed that the time delays between the images of the SN should depend on the distribution of matter in the lens and, geometrically, on the cosmic expansion rate. 
In \citet{kellyrodneytreu15}, we reported the first example of a strongly
lensed SN resolved into multiple images, which we found 
in near-infrared (NIR) {\it HST} WFC3 exposures of the MACS\,J1149+2223 cluster \citep{ebelingedgehenry01} 
taken as part of the Grism Lens-Amplified Survey from Space
(GLASS; PI T. Treu; GO-13459; see  \citealt{schmidttreubrammer14,treuschmidtbrammer15}). 
The data revealed a total of four images of the SN in an Einstein cross surrounding
an early-type galaxy in the cluster.
Here we use photometry and spectroscopy from the first year after discovery to classify the SN and characterize its basic properties.

The explosion site of SN Refsdal is close to the tip of a spiral
arm of a galaxy at redshift $z=1.49$ \citep{smithebelinglimousin09}.
The galaxy is inclined at an angle of $i=45\pm10^\circ$ \citep{yuankewleyswinbank11} and is multiply imaged \citep{zitrinbroadhurst09} by the potential of the massive MACS\,J1149+2223 cluster [($1.4 \pm 0.3) \times 10^{15}$ M$_{\odot}$; \citealt{vdlallen14,kellyvonderlinden14,applegatevdl14}] at $z=0.54$.
The cluster lens forms three images of the SN host galaxy that include the explosion site of the SN. 

Light that travels toward us on a direct route through the center of the cluster arrives last owing to the
greater spatial curvature and gravitational time dilation near the center of the potential (see \citealt{treuellis15} for a review). 
In \citet{kellyrodneytreu15}, we predicted that the 2014 appearances were only the next-to-last arrival and that the SN would reappear within several years closer to the center of the cluster. 
Modeling efforts \citep{oguri15, sharonjohnson15,diegobroadhurstchen15} 
sought to make more precise predictions by collecting improved datasets \citep{jauzacrichardlimousin15,treubrammerdiego15,grillokarmansuyu15,kawabataoguriishigaki15}.
An imaging campaign with {\it HST} (PI P. Kelly; GO-14199) detected the predicted reappearance on December 11, 2015 \citep[UT dates are used throughout this paper;][]{kellyreappear15}, and deep follow-up images will measure the relative time delay with 1--2\% precision.

\begin{deluxetable*}{cccccccccccc}
\tabletypesize{\scriptsize}
\tablecaption{Details of Keck-I MOSFIRE Observations
\label{tab:obskeck}}
\tablecolumns{11} 
\tablehead{
	\colhead {Date}&
	\colhead {$F160W$ AB}&
	\colhead {$\alpha$(J2000)}& 
	\colhead {$\delta$(J2000)}& 
	\colhead {Position}& 
	\colhead {Para.}& 
	\colhead {Airmass}& 
	\colhead {Total}&
	\colhead {Slit}&
	\colhead {Seeing}&
	\\
	\colhead {(MJD)} & 
	\colhead {($\pm$0.1 mag)} & 
	\colhead {} &
	\colhead {} &
	\colhead {Angle}& 
	\colhead {Angle}& 
	\colhead {}& 
	\colhead {Exp. (s)}& 
	\colhead {Width}& 
	\colhead {}& 
 } 
\startdata
56984 & $\sim25.45$ (S1); $\sim25.5$ (S2)  & 11:49:35.574 & +22:23:44.06 &
109.68$^{\circ}$ & 82$^{\circ}$  & 1.56  & 3578.8 & 0\farcs7 &
0\farcs7$^{(a)}$
\enddata
\tablecomments{ Magnitudes listed are total and do not account for slit losses, and are extrapolated from
the first $F160W$ observations by $\sim6$ days in the observed frame ($\sim2.5$ days in the
rest frame). Observations were acquired in a sequence of thirty 119.29\,s
exposures.
 $^{(a)}$Determined from the $J$-band image; the FWHM in the 
$H$ band should be smaller.}
\end{deluxetable*}

\begin{deluxetable*}{ccccccccccc}
\tabletypesize{\scriptsize}
\tablecaption{Details of VLT X-shooter Observations
\label{tab:obsxshooter}}
\tablecolumns{10} 
\tablehead{
	\colhead {OB}& 
	\colhead {Date}& 
	\colhead {$F160W$ AB}&
	\colhead {$\alpha$(J2000)}& 
	\colhead {$\delta$(J2000)}& 
	\colhead {Position}& 
	\colhead {Para.}& 
	\colhead {Airmass}& 
	\colhead {Total}&
	\colhead {NIR Slit}
	\\
	\colhead {}& 
	\colhead {(MJD)} & 
	\colhead {($\pm 0.1$ mag)} & 
	\colhead {} &
	\colhead {} &
	\colhead {Angle}& 
	\colhead {Angle}& 
	\colhead {}& 
	\colhead {Exp. (s)}& 
	\colhead {Width}
 } 
\startdata
1 & 57158 & 24.75 (S1); 24.47 (S2)&  11:49:35.520  & +22:23:44.44 &
108.2$^{\circ}$ & 175$^{\circ}$   & 1.47  & 4800 & 0\farcs6
 \\
2 & 57188 & 24.75 (S1); 24.60 (S2) & 11:49:35.511 & +22:23:44.65  &
108.2$^{\circ}$  &  165$^{\circ}$  & 1.51 & 4800 &  0\farcs9
\\
3 & 57189 & 24.60 (S2); 24.60 (S3) &  11:49:35.406 & +22:23:44.40  &
51.6$^{\circ}$  & 162$^{\circ}$  & 1.53 & 4800 &  0\farcs9

\enddata
\tablecomments{ Magnitudes listed are total and do not account for slit losses. Observations were
acquired in a sequence of four 1200\,s exposures. }
\end{deluxetable*}

\begin{deluxetable*}{ccc}
\tabletypesize{\scriptsize}
\tablecaption{Seeing During X-shooter Observation Blocks
\label{tab:seeing}}
\tablecolumns{3} 
\tablehead{
	\colhead {OB} & 
	\colhead {Acquisition Image} & 
	\colhead {Estimated Average} 
	\\
	\colhead{} &
	\multicolumn{2}{c}{(FWHM)} 
 } 
\startdata
1 &  0\farcs69 & 1\farcs00   \\
2 &  0\farcs62 & 0\farcs7  \\
3 &  0\farcs54 & 0\farcs9
\enddata
\tablecomments{ To estimate the seeing at the beginning of each OB, two bright stars in the acquisition image are fit with a two-dimensional
Gaussian. For all observations a
significant worsening of the seeing occurred during the observations. To estimate the average seeing during each OB, we scale the average DIMM seeing by the difference between the measured FWHM in the acquisition images and the DIMM FWHM recorded at the beginning of the OB. }
\end{deluxetable*}

Either a thermonuclear Type Ia SN or a core-collapse SN provided a reasonable 
match to the light curve and colors of SN Refsdal during the first month after
discovery, which was made on November 11, 2014.
After a 1\,hr Keck-I MOSFIRE observation (PI C. Steidel) was not able to detect the SN,  
an {\it HST} Director's Discretionary (DD) time program was carried out from \grism~(PI P. Kelly; GO-14041) to acquire
WFC3 spectra.
Instead of fading as would have been expected for a SN Ia, 
SN Refsdal continued a slow rise in brightness, 
which made ground-based NIR spectroscopy
possible near the peak of the light curve.
We obtained Very Large Telescope (VLT) X-shooter spectra through an ESO DD program (PI J. Hjorth;
295.D-5014) in May and June 2015, approximately six months after discovery.
Keck-II DEIMOS observations taken in December 2015, March 2015, and May 2015 
yielded no detection of the SN at optical wavelengths.

Here we show that the spectra and light curve of SN Refsdal are consistent with those
of SN~1987A-like supernovae (SNe), whose prototype was the best-studied SN explosion in 
recent history.
A companion paper \citep{rodneystrolgerkelly15} presents measurements of the relative time delays
and magnifications of the four images in the Einstein cross, and 
magnitudes measured using point-spread-function (PSF) fitting photometry. 
In Section~\ref{sec:data}, we describe the MOSFIRE, DEIMOS, {\it HST} grism, and
X-shooter spectra that we have collected. 
The photometric classification of the light curve is discussed in Section~\ref{sec:PhotometricClassification}.
Section~\ref{sec:snspec} presents an analysis of the SN spectra, and
Section~\ref{sec:snenvironment} contains measurements of the host-galaxy
environment.  
We summarize the results in Section~\ref{sec:summary}. 
The methods that we use to process and extract spectra of SN Refsdal are
explained in detail in the Appendix. 

\section{Data}
\label{sec:data}

 \begin{figure*}
 \begin{center}
 \includegraphics[width=6.5in]{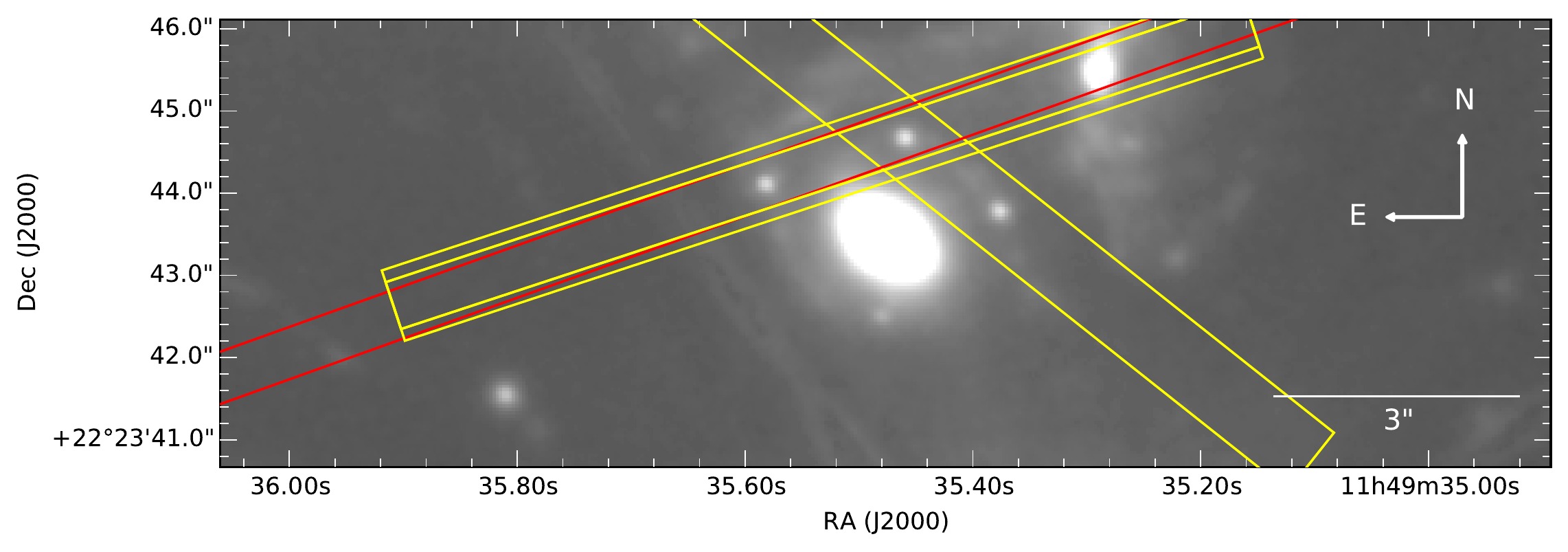}
 \caption{Image of the slit positions used for the MOSFIRE (red) and X-shooter (yellow) observations. The image is a coaddition of $F125W$ exposures which shows the four images forming the Einstein cross. }
\label{fig:slits}
\end{center}
\end{figure*}

\subsection{Keck-I MOSFIRE Spectra}
On November 23, 2014, approximately two weeks after discovery, 
we obtained a 1\,hr {\it H}-band integration with the Multi-Object Spectrometer For Infra-Red
Exploration (MOSFIRE; \citealt{mcleansteidelepps10,mcleansteidelepps12}) mounted
on the 10\,m Keck-I telescope.  As we show in Figure~\ref{fig:slits}, 
a 0\farcs7 wide slit was oriented position angle (PA) 109.68$^{\circ}$ 
to place the slit across both images S1 and S2.
The resolving power of the setup was $R = 3660$ in the $H$ band (1.48--1.81
$\mu$m), chosen to be able to have sensitivity to H$\alpha$ emission at the 
redshift (1.49) of the 
face-on spiral galaxy.  
Table~\ref{tab:obskeck} provides an overview of the MOSFIRE observations.

The data were obtained using mask nodding in thirty 119.29\,s exposures, split
evenly between integration at positions A and B separated by 12$''$ along the
slit in an alternating sequence, for a total integration of 3578.8\,s. 
The full width at half-maximum intensity (FWHM) of the $J$-band 
PSF was estimated to be 0\farcs7
during the observations, 
and the $H$-band FWHM is expected to have smaller size. 
The data were reduced using the MOSFIRE Data Reduction
Pipeline\footnote{https://keck-datareductionpipelines.github.io/MosfireDRP/}.

We extract the spectra of images S1 and S2 using a 4~pixel (0\farcs72) width
aperture centered on the expected positions of the images of the SN. 
The locations of S1 and S2 show narrow nebular emission from the host galaxy. 

\subsection{Keck-II DEIMOS Data}
We also obtained optical spectra of the field at the 10\,m Keck-II telescope using the DEIMOS spectrograph \citep{Faber03}. We used multislit masks with 1$''$ wide slits that included the positions of Refsdal images S1 and S3 on December 20, 2014 with approximately 3\,hr of exposure time in fair conditions, and images S2 and S3 on March 18, 2015 (3\,hr of exposure in good conditions) and May 20, 2015 (1.5\,hr of exposure in moderate conditions). We used the 600-line grating with a central wavelength of 7200\,\AA, resulting in a wavelength range of 5000--10,000\,\AA\ at a scale of 0.65\,\AA\ per pixel. A preliminary analysis of the data using a customized version of the DEEP2 pipeline \citep{Newman12,Cooper12} showed no detectable signal from SN Refsdal over the background.

\subsection{{\it HST} G141 Grism Spectra}
As a part of an {\it HST} DD program (GO-14041; PI P.
Kelly), we obtained thirty orbits of WFC3 G141 grism spectra during the period from
\grismstart\ through \grismend\ (13.2 days; 5.3 days in the rest frame) when SN
Refsdal had $F160W \approx 25.1$\,mag AB. During the grism observations, the mean phase
of the SN relative to maximum brightness was \grismphase.
Each 2405.9\,s G141 grism integration was followed by a WFC3 direct imaging
202.9\,s integration through either the $F125W$ or the $F160W$ broad-band
filters, 
which are used to align the grism data.
The total integration was split equally between observations at telescope orientations of
\one\ and \two, where the spectrum of the SN images S2 and S3 were expected to suffer the least contamination from spectra of nearby sources based on our planning simulations and knowledge of the layout of sources in the field.

The WFC3 G141 grism has a resolving power of $\sim70$\,\AA\ ($\sim1400$\,km\,s$^{-1}$), 
and a wavelength range of 11,000--17,000\,\AA\ ($\sim4400$--6800\,\AA\ in the
$z=1.49$ SN rest frame). Pointings were made using subpixel offsets to sample
the WFC3 PSF completely. 
The first order of the WFC3 G141 grism has a maximum efficiency of 48\% near
14,500\,\AA, while the second order reaches $\sim8$\% near $\sim$ 11,000\,\AA. 
Each WFC3-IR image has $1024 \times 1024$ pixels 
covering a $136'' \times 123''$ field of view, and
the spectra have an average tilt of $\sim0.5^{\circ}$ relative to the
detector's axis.

\subsection{{\it HST} Light Curves}
The extraction of the SN light curve from the {\it HST} imaging is described in a companion paper
\citep{rodneystrolgerkelly15}.

\subsection{VLT X-shooter Spectra}
Observations were acquired with the X-shooter echelle spectrograph
\citep{vernetdekkerdodorico11} mounted on Unit Telescope 2 (UT2) of the VLT in three observation
blocks (OBs) executed on May 16 (OB1), June 15 (OB2), and June 16 (OB3), 2015. 
X-shooter covers the entire spectral range 3100--25,000\,\AA\ by directing incoming light simultaneously to three arms with complementary wavelength coverage.
The observations of SN Refsdal were taken in nodding mode where positions A and B were separated by $7''$ along the slit.

The fraction of light from a well-centered point source that enters a spectrograph slit depends on the 
slit width and the FWHM of the PSF. 
From an $R$-band acquisition image, we can directly estimate the PSF FWHM through X-shooter at the beginning of each OB. 
The European Southern Observatory (ESO) Ambient Conditions
Database\footnote{http://archive.eso.org/cms/eso-data/ambient-conditions.html}
archives an estimate of the seeing  at Cerro Paranal from the differential image motion monitor (DIMM).
However, the DIMM seeing differs, in general, from the seeing achieved through X-shooter.
The DIMM shows that conditions changed significantly during each of the three OBs.

In Table~\ref{tab:seeing}, we list estimates for the average seeing during each OB.
We measure the seeing through X-shooter at the beginning of the OB from the FWHM of 
stars in the acquisition image.  
We then find the average FWHM recorded by the DIMM during the entire OB, 
and rescale this average value by the ratio between the  
DIMM FWHM at the start of the OB and the FWHM measured from the X-shooter acquisition image.
The DIMM PSF measured a degradation from a FWHM of $\sim 0\farcs8$ to $\sim 1\farcs6$ and then
settling at $\sim 1\farcs4$
over the course of the OB1 observations. 
During the OB2 observation, the seeing gradually improved from $\sim 0\farcs75$ to
$\sim 0\farcs65$.
During the OB3 observation, the seeing evolved from $\sim 0\farcs7$ to
$\sim 0\farcs9$ to $\sim 0\farcs8$.

An overview of the observations is given in Table \ref{tab:obsxshooter}, and we show the slit positions in Figure \ref{fig:slits}.
Observations of SN Refsdal were acquired at a high airmass almost orthogonal to the parallactic angle, so we need to consider carefully the effects of atmospheric dispersion  \citep{filippenko82}. 
The telescope tracks the target in images taken in 4700\,\AA, and 
a tip-tilt mirror corrects for the atmospheric refraction between 4700\,\AA\ and
the middle of the atmospheric dispersion range for the NIR arm at 13,100\,\AA.
Since the atmospheric dispersion in the NIR is comparatively small, X-shooter does not have
an atmospheric dispersion corrector (ADC) to correct the NIR arm.  
The relative shift between 13,100\,\AA\ and H$\alpha$ ($\sim$ 16,330\,\AA) is expected to be only $\sim0\farcs1$.
The relative atmospheric dispersion across the visible (VIS) arm is greater, and the ADC is not operational.  
We expect a $\sim 0\farcs5$ shift at [\ionpat{O}{ii}] ($\sim9274$\,\AA) relative to 4700\,\AA, the tracking wavelength where the target is centered on the slit. The slit width in the VIS spectroscopic arm is 0\farcs9 for OB1 and 1\farcs2 for OB2 and OB3. 
We test that the emission-line ratios in the NIR arm are not significantly affected by
comparing spectra from OB1 and OB2 to OB3, for which the PA was closer to the parallactic angle.

\begin{figure*}
\centering
\subfigure{\includegraphics[angle=0,width=6.5in]{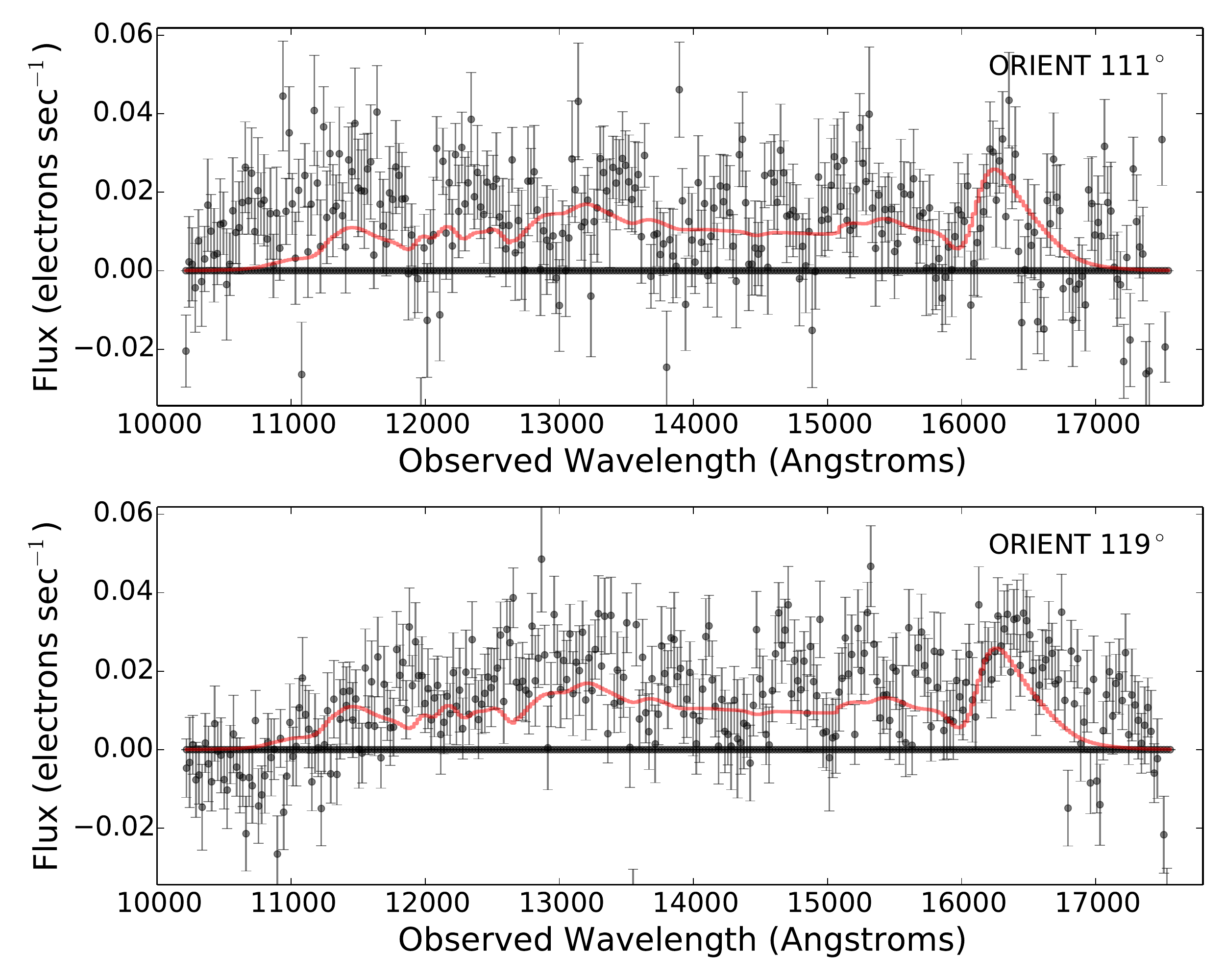}}
\caption{Spectra of image S2 extracted from grism data taken in the \one~and
\two~{\it HST} orientations. The spectra taken in a pair of 
different orientations of the telescope, selected to minimize overlap with the
traces of other objects, make it possible to obtain spectra that
should contain different residuals from removal of overlapping traces.  The
superimposed spectrum of SN 1987A is 
normalized to the average $F160W$ magnitude of S2 during
the period of observations. }
\label{fig:orientsoned}
\end{figure*}

 \begin{figure*}
 \begin{center}
 \includegraphics[width=6.5in]{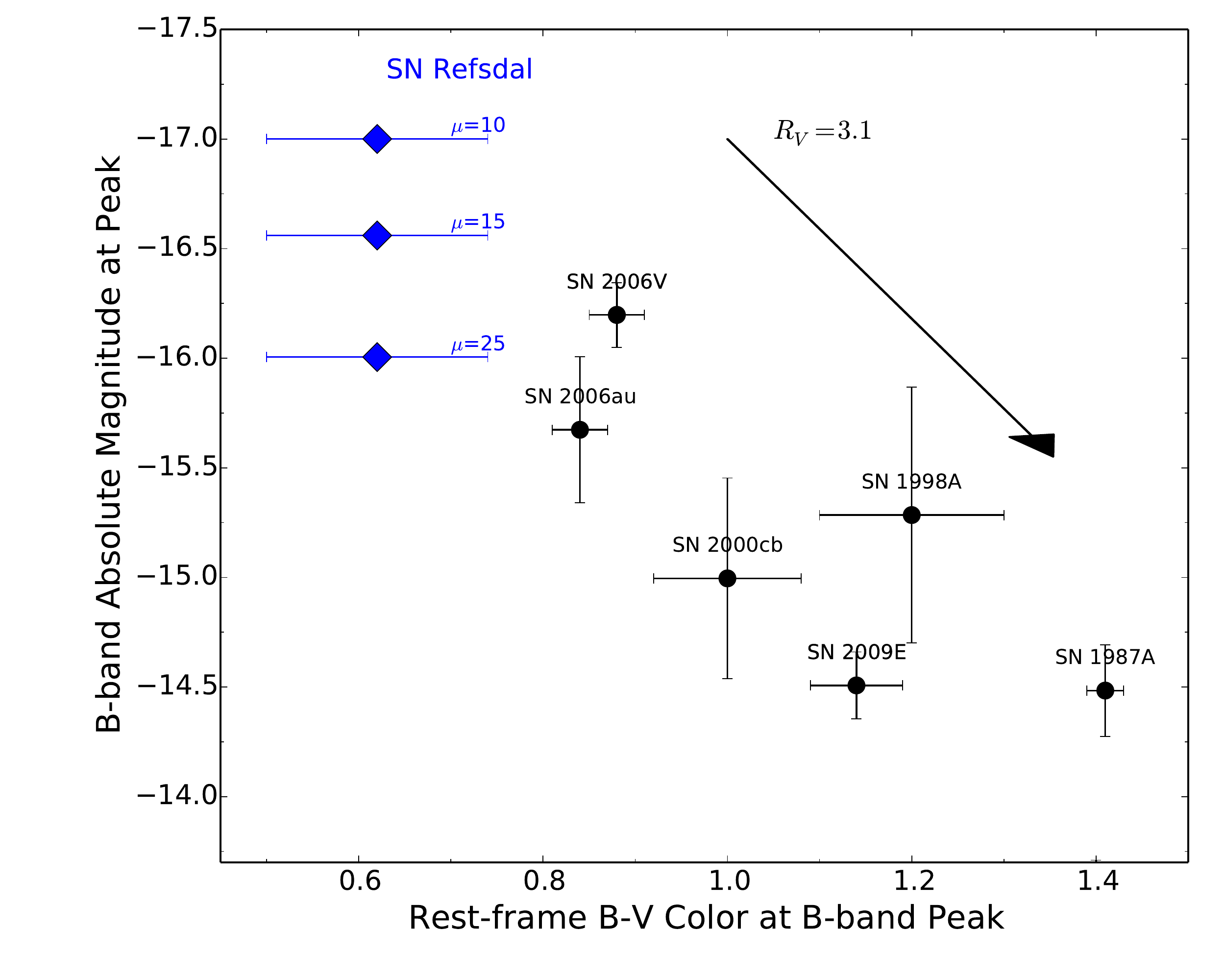}
 \caption{$B-V$ color and $B$-band absolute magnitude of SN Refsdal near light-curve peak in comparison to examples of low-redshift SN~1987A-like SNe.  If a SN~1987A-like event, SN Refsdal would be the most blue ($B-V$, although not $V-R$) and potentially have the most luminous $M_B$ absolute magnitude near the peak of its light curve, depending on the magnification.  As seen in Figure~\ref{fig:eightysevenalightcurve}, the slowly evolving light of SN Refsdal may be best matched by that of NOOS-005, which had the most luminous $M_I$ absolute magnitude near peak but lacked photometry through other broadband filters or spectroscopy.  The $B-V$ color and $B$-band luminosity of the SN~1987A-like SNe are corrected for reddening and extinction along the line of sight, but no correction is applied to the photometry of SN Refsdal.  For illustration, we plot a reddening vector for $E(B-V)=0.33$ mag for an $R_V = 3.1$ extinction law. Table~\ref{tab:magnifications} lists the magnifications of the images of the SN predicted by models of the galaxy and cluster lenses. }
\label{fig:absmagcolor}
\end{center}
\end{figure*}

\section{Photometric Classification}
\label{sec:PhotometricClassification}

\begin{figure*}
\centering
\includegraphics[width=\textwidth]{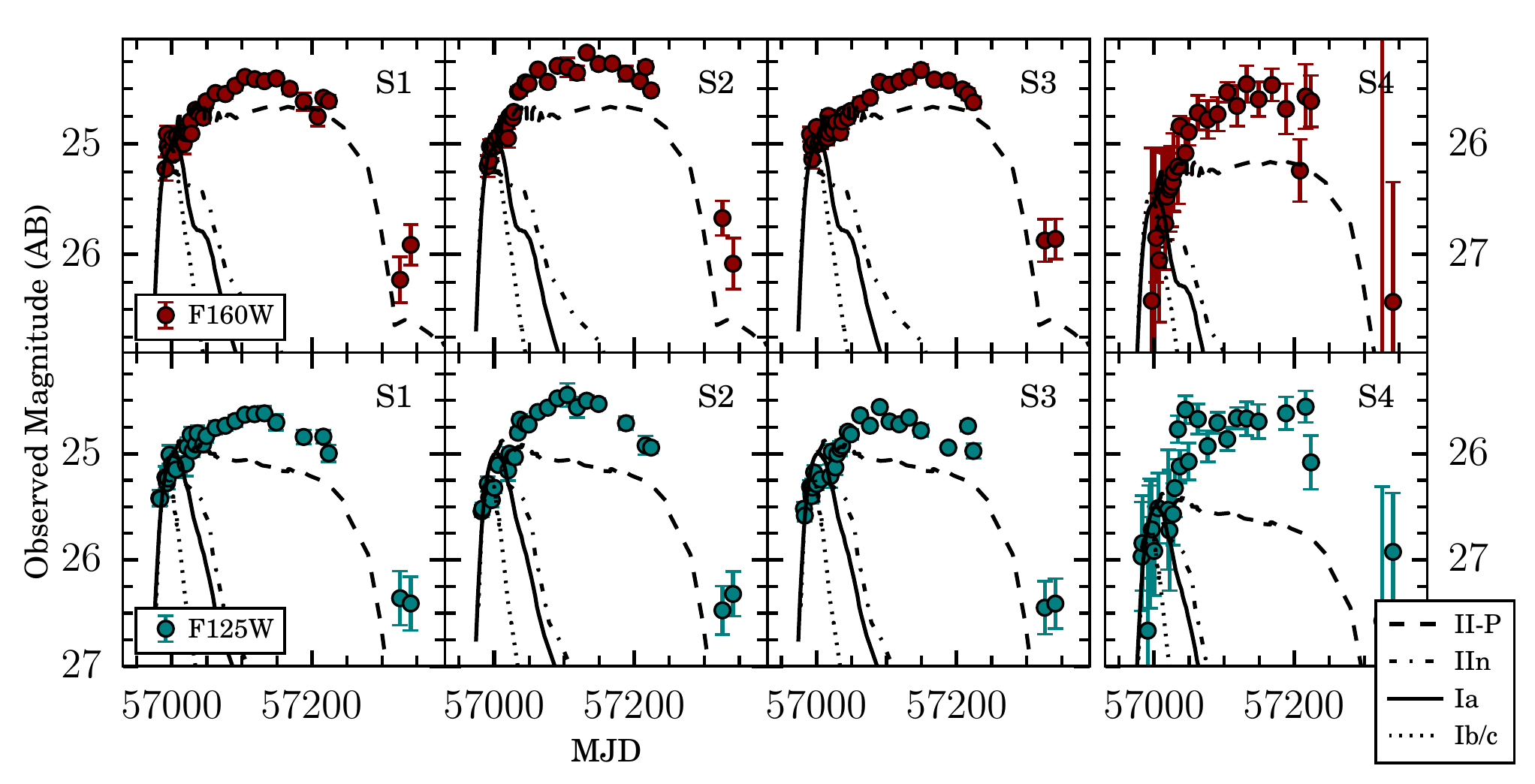}
\caption{Four columns of panels show light curves of images S1--S4 of SN Refsdal from left to right.  Top and bottom rows show the $F160W$ and $F125W$ photometry, respectively.  The SN~Ia and SN~Ib/c templates are clearly incompatible with the broad light-curve shape of SN Refsdal, while normal SNe~IIP (``plateau'') do not show rising luminosities during their plateau phase.}
\label{fig:LightcurveClassification}
\end{figure*}

During the first three months of observations after discovery on November 11, 2014, SN Refsdal continued to rise in
brightness well beyond the time where any normal Type Ia, Ib, or Ic SN
would have reached its peak luminosity \citep{kellyrodneytreu15}.  Continued monitoring showed that the
light curve rose for $\sim150$ days, and the SN became similarly
incompatible with most normal Type II SN light curves,
which brighten to a ``luminosity plateau'' over only several days to weeks in the rest frame
\citep{barbonciattirosino79,doggettbranch85}. 
In Figure~\ref{fig:LightcurveClassification}, we plot a comparison between the full SN light curves for S1--S4 against SN light-curve models for typical SNe~Ia, SNe~Ib/c, and SNe~II. 

This slow rise in brightness to a broad peak is consistent with the light curve of
SN 1987A, a peculiar Type II SN in the Large Magellanic Cloud (LMC) that is 
the nearest and brightest SN observed in the last four centuries
\citep[][and references therein]{arnettbahcallkirshner89}. 
SN 1987A brightened steadily in the rest-frame optical after initial emission from the shock breakout subsided to reach 
peak luminosity $\sim84$ days after first light; see Figure 3 from
\citet{fil97} for a comparison with SNe~IIP.   
The progenitor of SN 1987A was identified as a blue supergiant (Sk $-69^{\circ}202$; \citealt{gilmozzicassatellaclavel87,sonnebornaltnerkirshner87}) in the LMC ($d \approx 50$\,kpc), and more recent SNe with similar light-curve shapes and spectra are understood to be the result of the explosions of these compact massive stars.  
Well-studied examples of nearby SN~1987A-like events include SN 1998A
\citep{pastorellobaronbranch05}, SN 2000cb and SN 2005ci
\citep{kleiserpoznanskikasen11}, SN 2006V and SN 2006au
\citep{taddiastritzingersollerman12}, and SN 2009E
\citep{pastorellopumonavasardyan12}. 
These show a wide range of light-curve shapes and
colors, explosion energies, and expansion velocities
\citep[e.g.,][]{pastorellopumonavasardyan12,taddiastritzingersollerman12}.

\begin{figure*}
   \centering
   \includegraphics[width=\textwidth]{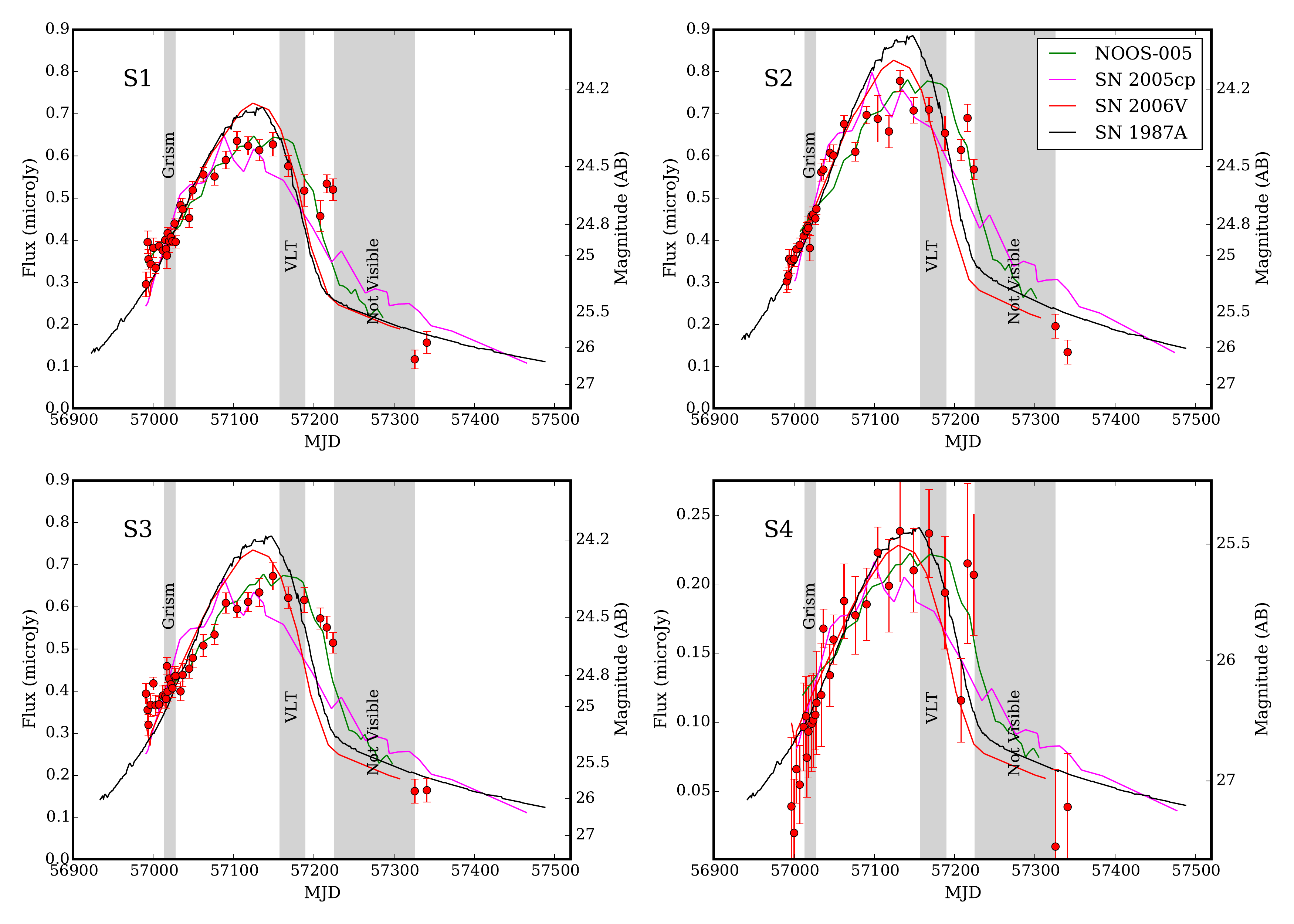}
   \caption{Comparison of SN Refsdal photometry to the light curves of SN~1987A-like SNe and the Type IIn SN 2005cp. NOOS-005 has the broadest peak and provides the best match to that of the SN Refsdal among examples of SNe with SN~1987A-like light curves. The color or spectroscopic properties of NOOS-005 are not known, because it was observed only through the $I$ band, and it reached a luminous absolute magnitude.  While SN 2005cp has a different spectroscopic classification which indicates the presence of significant CSM interaction, it may also be an explosion of a blue supergiant progenitor.  }
\label{fig:eightysevenalightcurve}
\end{figure*}

However, one published example of a SN~IIn, SN 2005cp \citep{kiewegalyamarcavi12}, shows a light curve similar to that of SN 1987A. SNe~IIn are characterized by relatively narrow
H emission lines originating in the interaction between the expanding ejecta and pre-existing 
CSM. The interaction produces narrow Balmer emission and can
power a strong underlying continuum.  
In Figure~\ref{fig:eightysevenalightcurve}, we compare the $F160W$ (rest-frame $\sim R$-band) light curves of the four SN Refsdal images with the $R$-band light curves of SN~1987A-like SNe (SN 2006V and NOOS-005, SN~1987A) and the Type IIn SN 2005c. 

\begin{figure}
\centering
\includegraphics[width=\columnwidth]{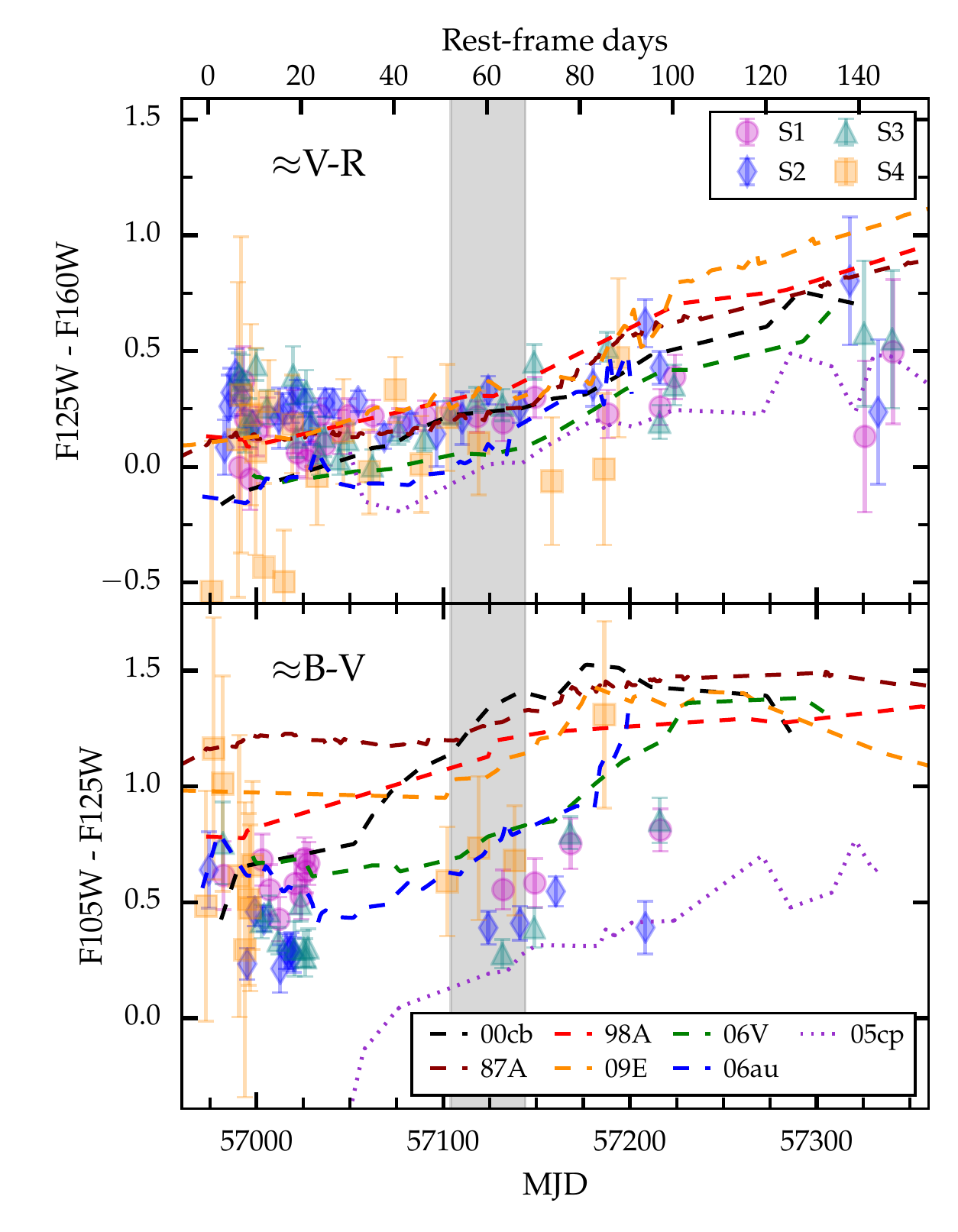}
\caption{Comparison of the $F105W-F125W$ and $F125W-F160W$ colors of SN Refsdal with those of SN~1987A-like SNe as well as the Type IIn SN 2005cp as a function of phase relative to maximum light. 
At all phases for which we have photometry, SN Refsdal shows a $F125W-F160W$ ($\sim V-R$) color consistent with those of SN 1987A-like SNe. 
At an early phase, SN Refsdal exhibits a $F105W-F125W$ ($\sim B-V$)  color that may be comparable to those of SN 2006V and SN 2006au, the bluest known example of a SN~1987A-like event. Near maximum light, SN Refsdal may be $B-V \approx 0.1$--0.2\,mag bluer than SN 2006V and SN 2006au.  The $B-V$ color of SN 2005cp, a SN~IIn whose light curve resembles that of SN 1987A, is bluer than that of SN Refsdal at all phases having photometry.  }
\label{fig:ColorCurveComparison}
\end{figure}

In Figure~\ref{fig:ColorCurveComparison}, we show that SN Refsdal has a $F125W-F160W$ ($\sim V-R$ in the rest frame of the SN) color comparable to those of nearby SN~1987A-like SNe.  At an early phase, SN Refsdal has a $F105W-F125W$ ($\sim B-V$) color comparable to those of SN 2006V and SN 2006au, the bluest SN~1987A-like SNe, but near maximum brightness the color of SN Refsdal is blue in comparison to even SN 2006V and SN 2006au. 
At almost all epochs and in both colors, SN 2005cp is bluer than SN Refsdal. Here we have applied K-corrections to the extinction-corrected colors of the comparison sample (see Table~\ref{tab:comparisondata}). We apply no correction for possible dust extinction to the colors of SN Refsdal, since the low signal-to-noise ratio (S/N) or spectral resolution of the spectra do not allow any constraint on absorption by, for example, Na~I~D or diffuse interstellar bands.

For the sample of SN~1987A-like SNe included in \citet{pastorellopumonavasardyan12}, 
the SNe show a range of absolute magnitudes (in the $V$, $R$, and $I$ bands) from about $-15$
to $-17.5$\,mag. In Figure~\ref{fig:absmagcolor}, we show that SN Refsdal has a brighter $M_B$ absolute magnitude and a bluer $B-V$ color near maximum light than the comparison sample of SN~1987A-like SNe.  The magnifications predicted for sources S1, S2, and S3 are listed in Table~\ref{tab:magnifications}. For a peak magnitude of $F160W \approx 24.25$\,mag AB (see \citealt{rodneystrolgerkelly15}), SN Refsdal would have had an $M_R = -17.5$ mag (magnification $\mu = 25$) to $M_R = -18.5$ mag ($\mu = 10$). 
The range of possible absolute magnitudes imply that SN Refsdal would likely be the most luminous well-studied SN~1987A-like event.

In Figure~\ref{fig:eightysevenalightcurve}, the broad-peaked shape of the light curve of OGLE-2003-NOOS-005\footnote{http://ogle.astrouw.edu.pl/ogle3/ews/NOOS/2003/noos.html}
provides the best match to that of SN Refsdal. 
NOOS-005 was observed only in the $I$ band and no spectrum was taken, so its color was not measured, but
its $I$-band absolute magnitude ($M_I=-17.51$\,mag) was the most luminous among those of SN~1987A-like SNe.

\begin{figure*}
\centering
\subfigure{\includegraphics[angle=0,width=6.5in]{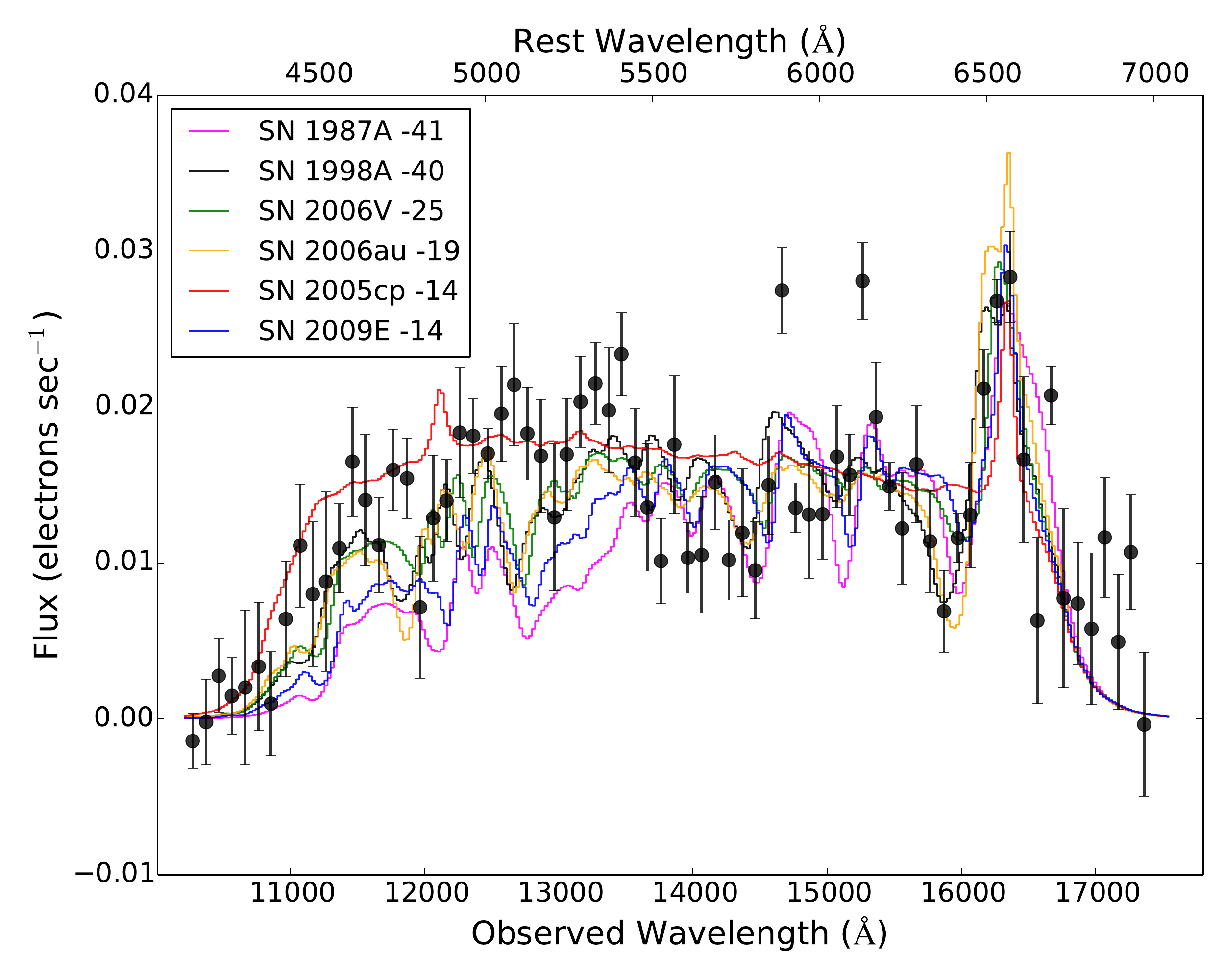}}
\caption{ Combination of grism spectra taken in both \one\ and \two\ orientations
and binned in wavelength.  Each bin is \binwidth\ in width, and plotted
uncertainties are estimated through bootstrapping with replacement fluxes in the
20\,\AA\ bins. The grism spectra contributing to this combined spectrum have
phases of \grismphase\ days. SN Refsdal exhibits stronger H$\alpha$ emission and a
bluer continuum at this phase. The color of the SN measured from coadded
direct images taken after each grism integration shows 
agreement with that computed from synthetic magnitudes.  }
\label{fig:combined}
\end{figure*}

\begin{figure*}
\centering
\subfigure{\includegraphics[angle=0,width=6.5in]{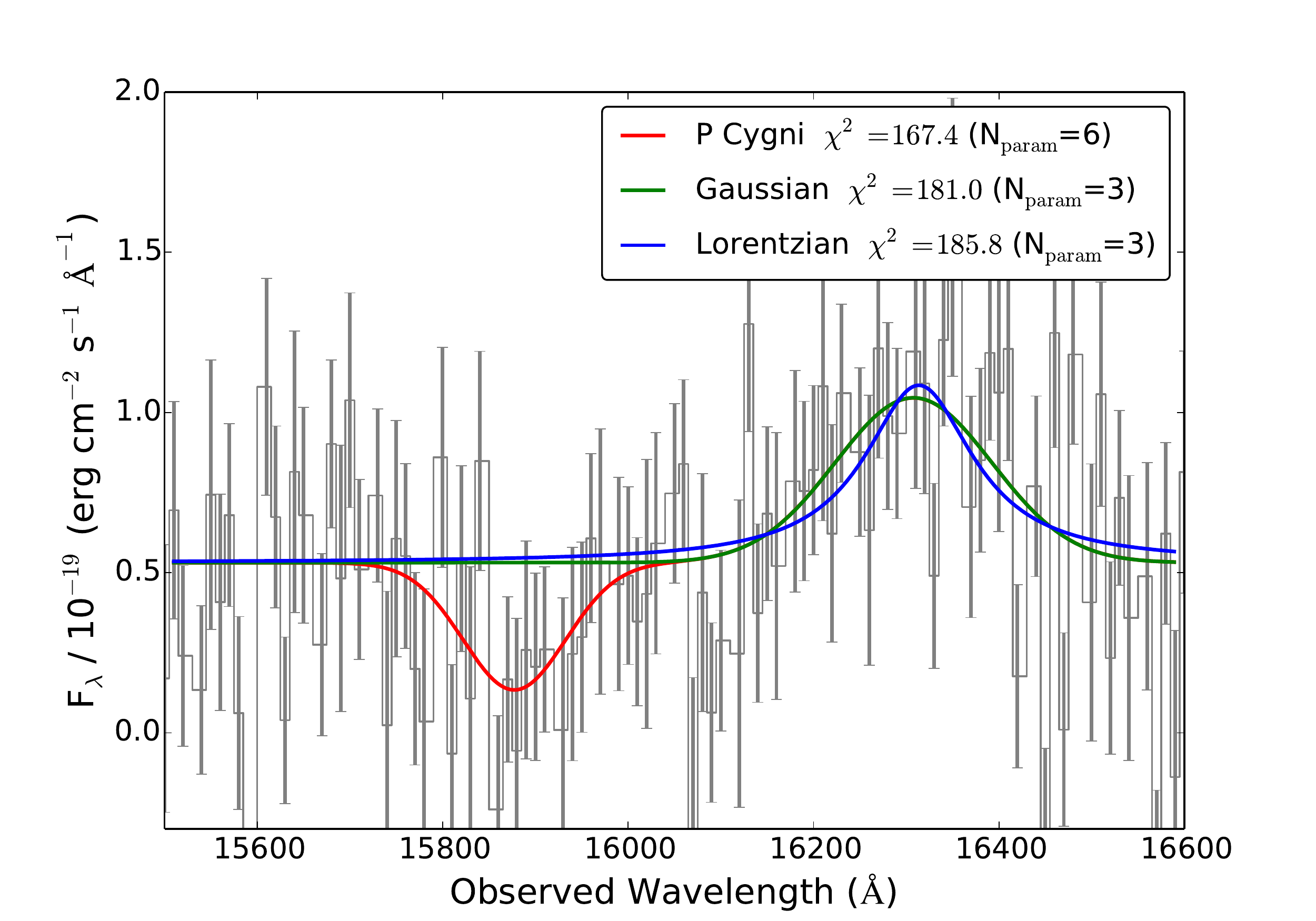}}
\caption{Model fits to {\it HST} WFC3 G141 grism spectrum of H$\alpha$
taken \grismphase\ relative to maximum light.
Table~\ref{tab:aic} lists the changes in the AIC between the model fits. 
While the $\chi^2$ values listed are from fitting the unbinned spectra, 
we plot the spectra after binning above.
 }
\label{fig:grismmodel}
\end{figure*}

\begin{figure*}
\centering
\subfigure{\includegraphics[angle=0,width=6.5in]{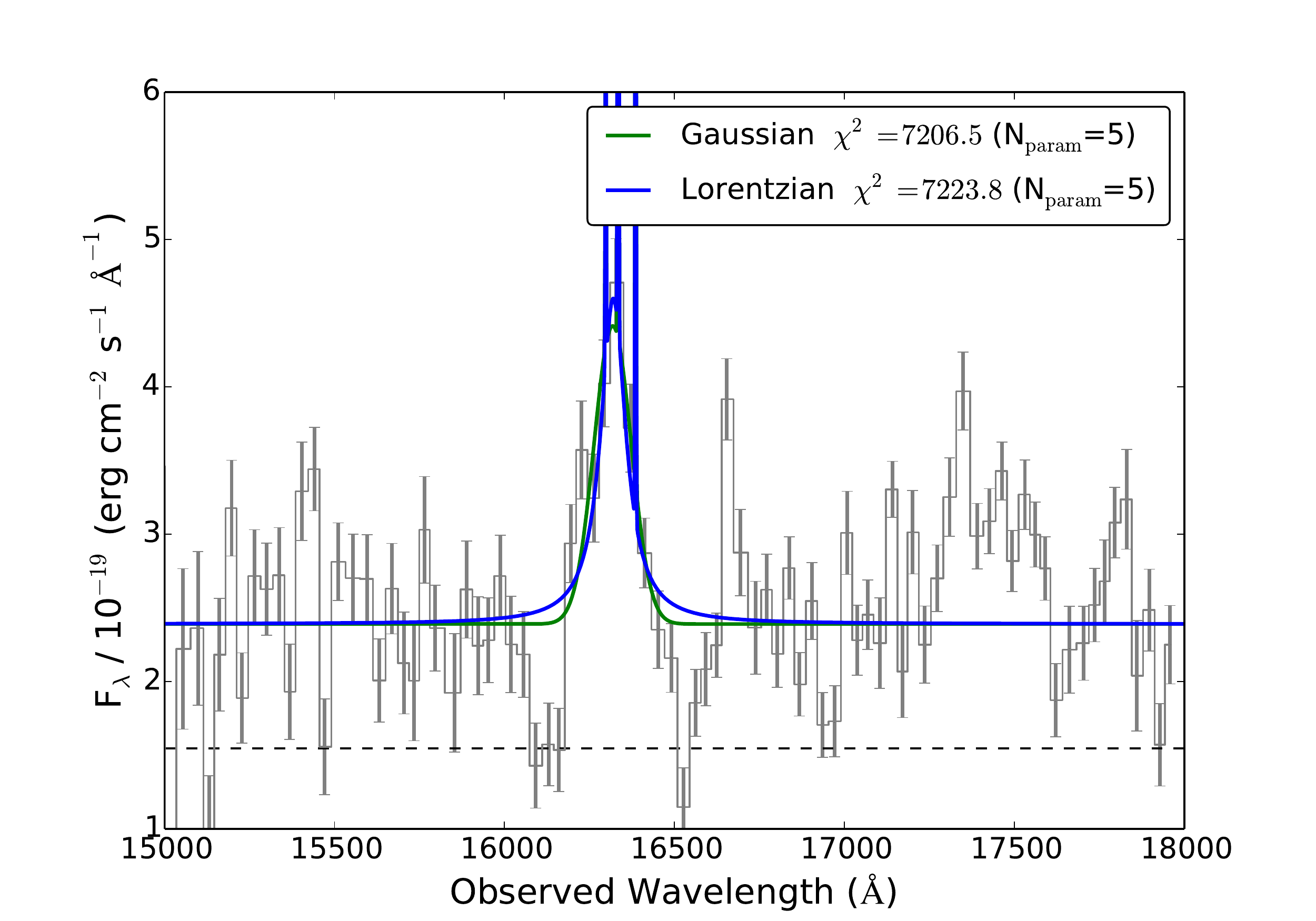}}
\caption{Model fits to the VLT X-shooter spectrum of H$\alpha$ taken
\xshooterphase\ relative to maximum light at the position of SN Refsdal.
Table~\ref{tab:aic} lists the changes in the AIC between the model fits. 
While the $\chi^2$ values listed are from fitting the unbinned spectra, 
we plot the spectra after binning above.
}
\label{fig:vltmodel}
\end{figure*}

\begin{figure*}
\centering
\subfigure{\includegraphics[angle=0,width=5.5in]{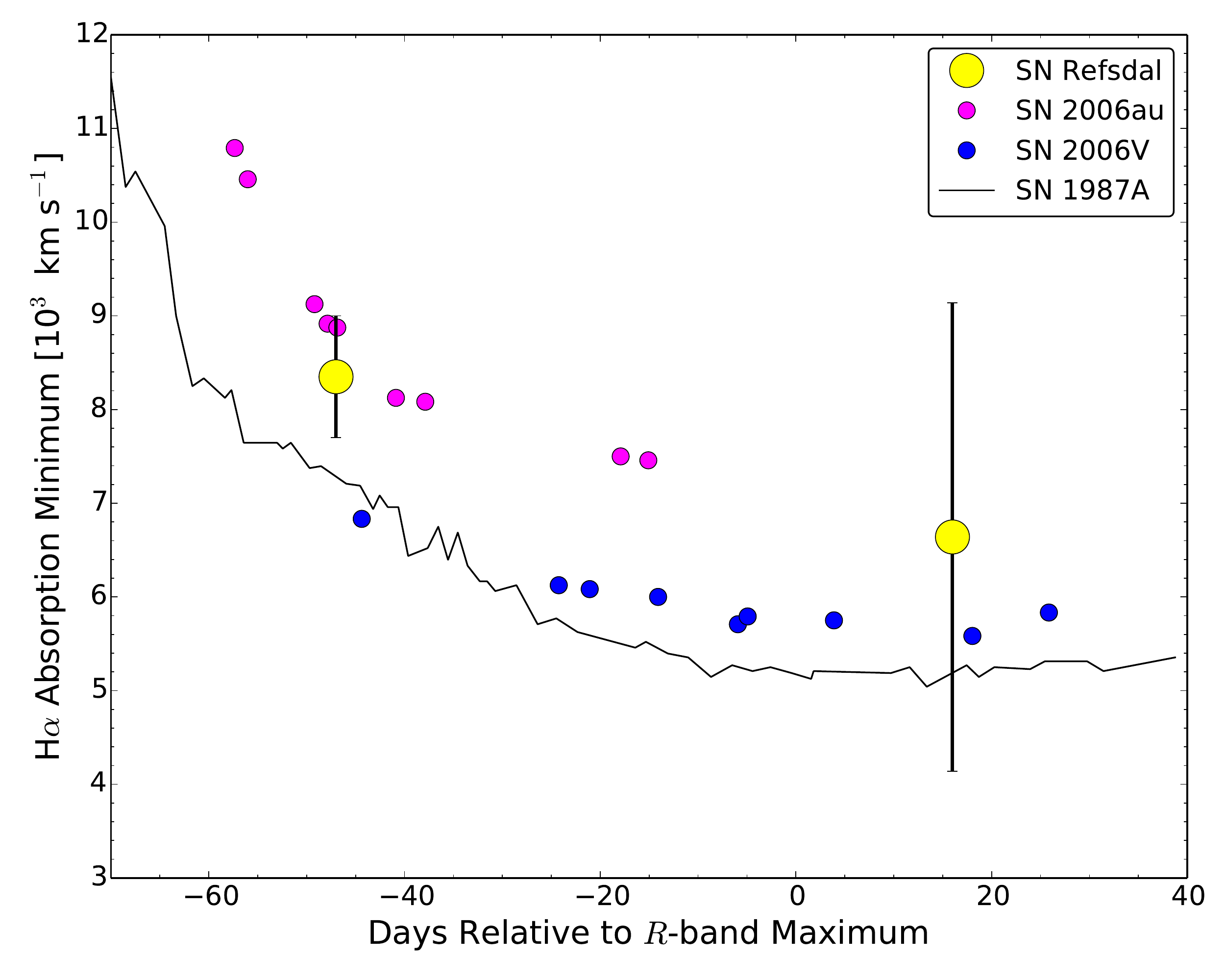}}
\caption{Comparison between constraints on H$\alpha$ P-Cygni absorption expansion velocity 
and H$\alpha$ expansion velocities of SN~1987A-like SNe.  The plotted measurements of SN 2006V, SN 2006au, and SN 1987A are from \citet{taddiastritzingersollerman12}.  The early-time H$\alpha$ expansion velocity of SN Refsdal is from the WFC3 grism spectra, and the later constraint is from the X-shooter spectra.
}
\label{fig:halphaminimum}
\end{figure*}

\begin{figure*}
\centering
\subfigure{\includegraphics[angle=0,width=6.5in]{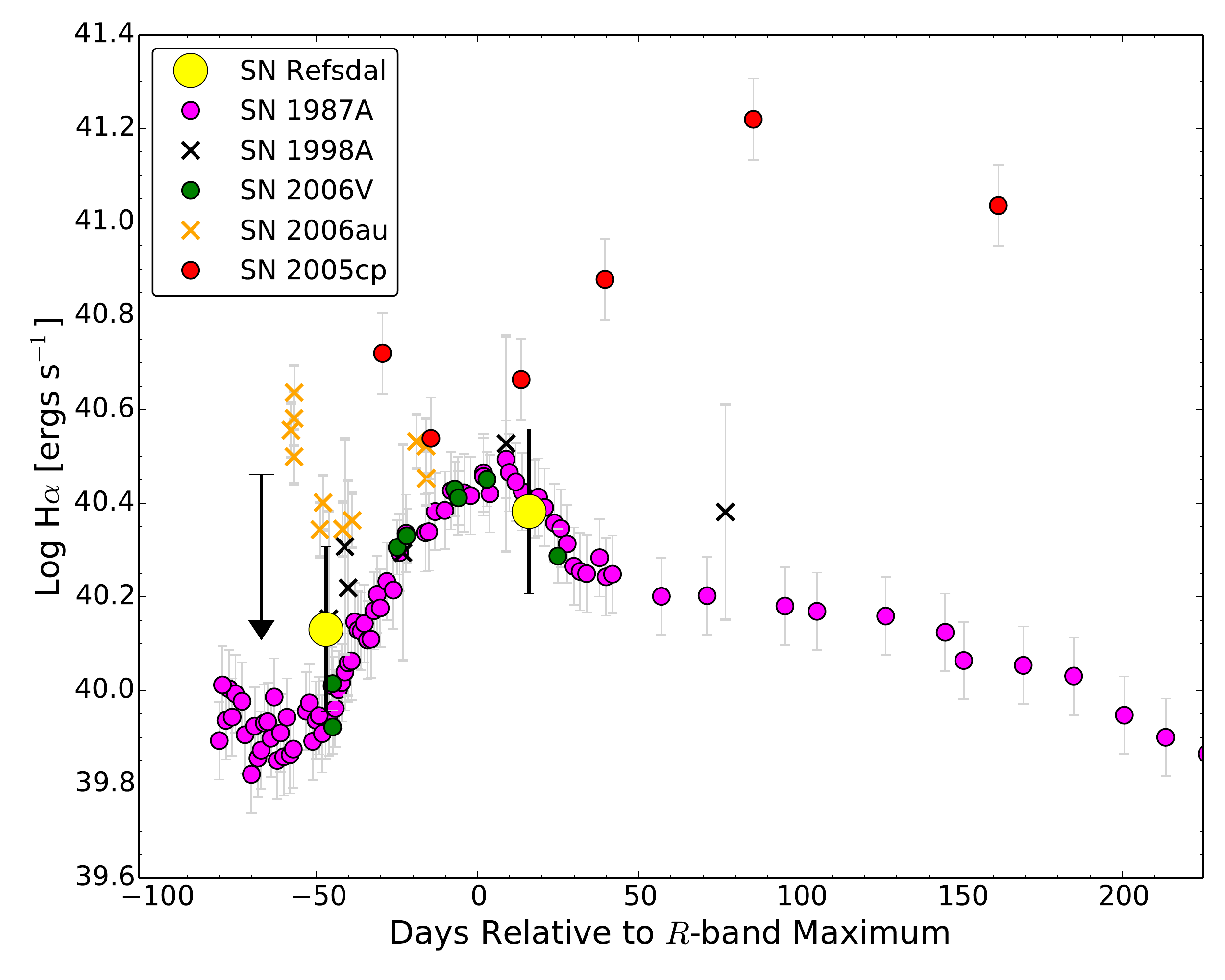}}
\caption{Comparison between the H$\alpha$ luminosity evolution of SN Refsdal (after correcting for magnification) and
that of SN with SN~1987A-like light curves. The SNe plotted for comparison were
spectroscopically classified as Type IIP, with the exception of SN 2005cp. The
relatively narrow Balmer emission lines of SN 2005cp led it to be classified as a SN~IIn,
although its light curve is similar to those of SN~1987A-like SNe. 
The upper limit on the H$\alpha$ luminosity obtained \mosfirephase\ before $R$-band maximum is from the MOSFIRE integration, the measurement at
\grismphase\ is from the WFC3 G141 grism data, and the measurement at
\xshooterphase\ is from the X-shooter spectrum. 
Plotted error bars correspond to uncertainties in the distance to the explosion,
or in the case of SN Refsdal, magnification from the cluster. 
}
\label{fig:halphastrength}
\end{figure*}

\begin{figure*}
\centering
\subfigure{\includegraphics[angle=0,width=6.5in]{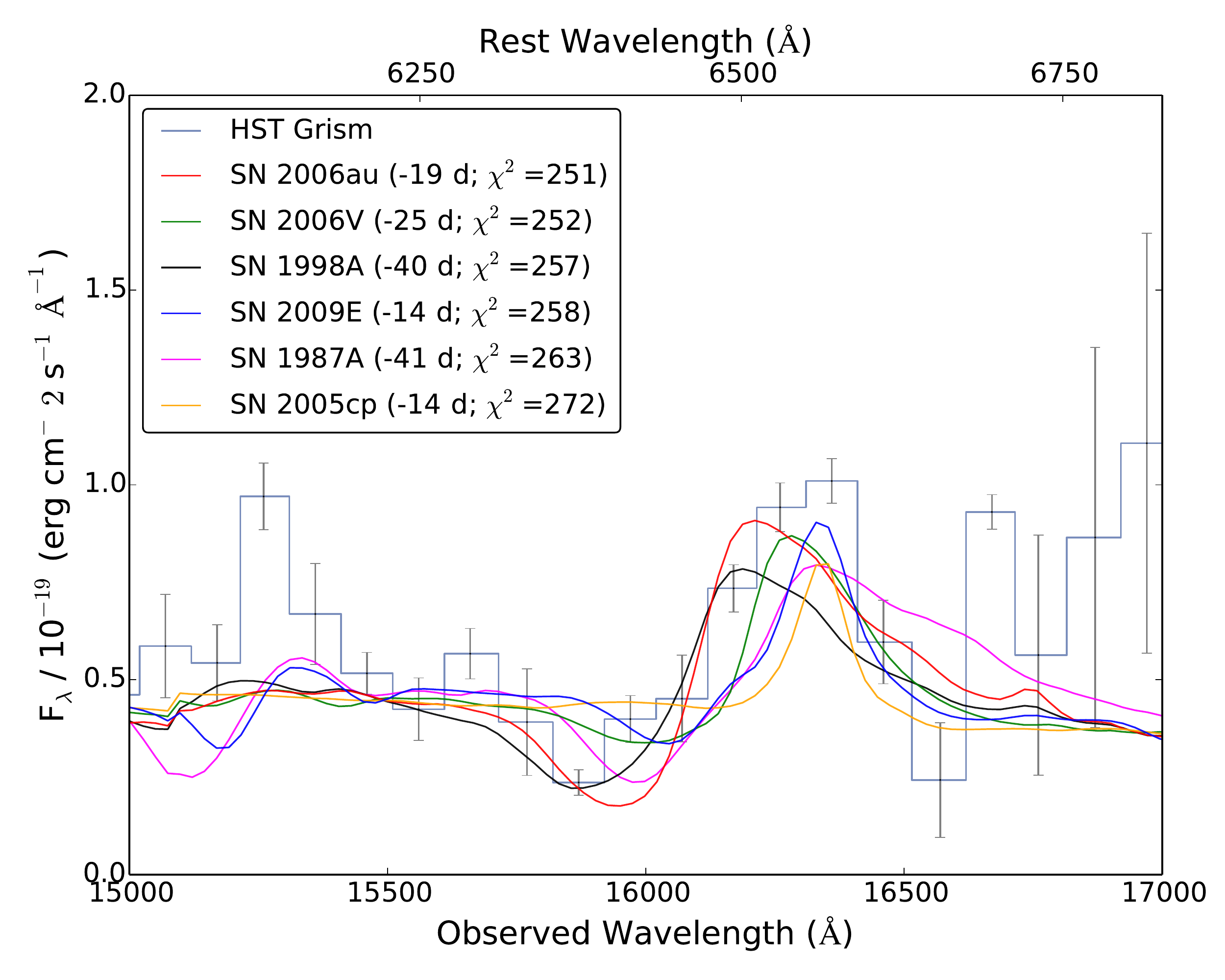}}
\caption{Comparison between the {\it HST} WFC3 G141 grism spectrum of H$\alpha$
taken \grismphase\ relative to maximum light and the spectra of SN~1987A-like SNe taken
at a roughly similar phase.
Here we limit the comparison to a spectral region near the broad H$\alpha$. 
We have used a simulation of the grism to compute the expected spectrum, and
these processed spectra are plotted. 
We compute the $\chi^2$ agreement between the grism spectrum and the comparison
spectra. 
Interpretation of $\chi^2$ values on the basis of Kullback-Leibler information
entropy \citep{aka74,sug78}
indicates that a $\chi^2$ difference of 2 is evidence against the more poorly fitting model and 
6 constitutes strong positive evidence (e.g., \citealt{kas95}; \citealt{muk98}). The grism
data disfavor the relatively narrow H$\alpha$ profiles of SN 2005cp, a SN~1987A-like
SN that was classified as a SN~IIn.
 }
\label{fig:grismsncomp}
\end{figure*}

\begin{figure*}
\centering
\subfigure{\includegraphics[angle=0,width=6.5in]{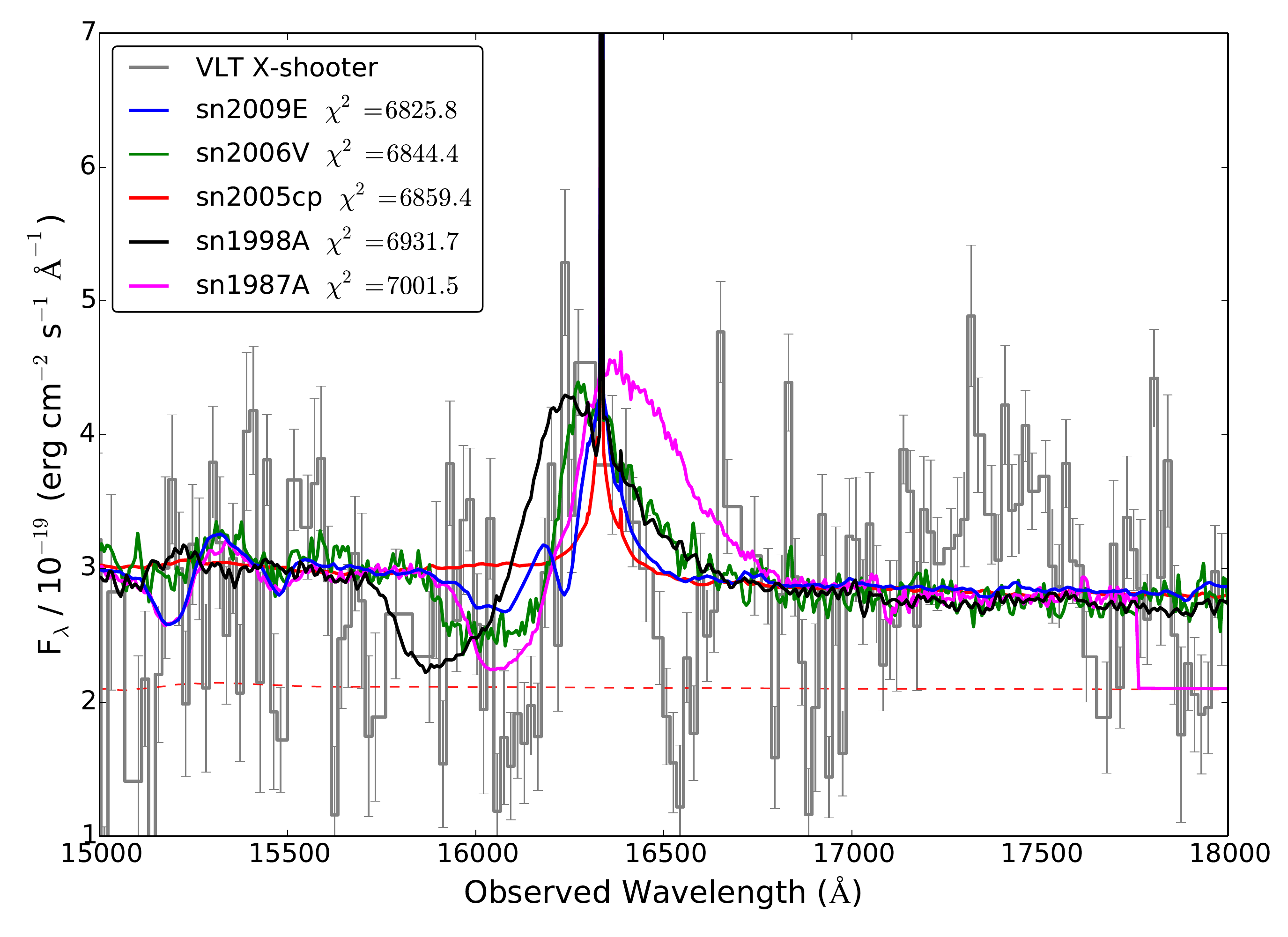}}
\caption{Comparison between the VLT X-shooter spectrum of H$\alpha$
taken \xshooterphase\ relative to maximum light and the spectra of SN~1987A-like SNe taken
at a roughly similar phase.
Here we limit the comparison to a spectral region near the broad H$\alpha$. 
We compute the $\chi^2$ agreement between the X-shooter spectrum and the comparison
spectra. Interpretation of $\chi^2$ values on the basis of Kullback-Leibler information
entropy \citep{aka74,sug78} indicates that a $\chi^2$ difference of 2 is evidence against the more poorly fitting model and 6 constitutes strong positive evidence (e.g., \citealt{kas95}, \citealt{muk98}).
The VLT data disfavor the relatively narrow H$\alpha$ profiles of SN 2005cp, a SN~1987A-like
SN that was classified as a SN~IIn.
 }
\label{fig:vltsncomp}
\end{figure*}

\section{Spectroscopic Classification and Characteristics}
\label{sec:snspec}

The light curve and colors of SN Refsdal can be matched approximately by those of SN~1987A-like SNe~II or, alternatively, SN 2005cp, a SN~IIn. We next use the WFC3 grism and the VLT X-shooter spectra to confirm the Type II classification spectroscopically, and find evidence from the H features that strongly favors the identification of SN Refsdal as a SN~1987A-like SN instead of a SN~IIn.

Figure~\ref{fig:combined} plots the binned WFC3 grism spectra of SN Refsdal taken in both the \one\ and \two\ telescope orientations. We plot the weighted
average of the raw flux measurements (each 20\,\AA) within each 
\binwidth~wavelength bin, and plot an uncertainty computed using bootstrapping
with replacement. 
We show, for comparison, a spectrum of SN 1987A
obtained at $-$41 days,
of SN 1998A at $-$40 days \citep{pastorellobaronbranch05},
of SN 2006V at $-$25 days, and of SN 2006au at $-$19 days
\citep{taddiastritzingersollerman12}.
These spectra are scaled so that $F160W = $~\magonesixty, the average flux of the SN during the grism observations.

In Figures~\ref{fig:grismmodel} and \ref{fig:vltmodel}, we plot the spectral region near 
H$\alpha$ for the WFC3 grism and X-shooter spectra.
These exhibit broad H$\alpha$ emission that spectroscopically classifies SN Refsdal as a Type II SN.

\subsection{H$\alpha$ P-Cygni Absorption}
A strong and wide H$\alpha$ P-Cygni absorption feature 
is present in the spectra SN~1987A-like SNe~II but absent from the spectra of SNe~IIn.
In SN~1987A-like SNe~II, the characteristic broad and deep P-Cygni absorption develops as the 
photosphere recedes into the ejecta. 
By constrast, the ejecta of SNe~IIn collide with CSM and the photosphere 
generally forms in proximity to the heated, shocked material.

First, we smooth the SN spectrum using a $\sigma=2000$\,km\,s$^{-1}$ Gaussian kernel and variance weighting. 
After removing $>5\sigma$ outliers from the data, we resmooth the spectrum with outliers removed using the same kernel.  We search the smoothed spectrum across the wavelength range 15,300--16,100\,\AA\ to find the absorption minimum.

To determine the uncertainty of the wavelength of the absorption minimum, we use boostrap resampling. 
For the grism and X-shooter spectra, we assemble the set of all flux measurements at each wavelength taken in both orientations or in all combinations of OBs and SN images, respectively.
We resample these sets of measurements with replacement to create the full set of bootstrapped spectra. 

The distribution of absorption minima we measure from the bootstrapped WFC3 grism spectra is approximately Gaussian. 
After rejecting a small population of $>5\sigma$ outliers, 
we find an absorption minimum of \mbox{\halphaabsorbminimumvelgrism$\pm$\halphaabsorbminimumvelstdgrism\,km\,s$^{-1}$} ({\mbox{\halphaabsorbminimumwavegrism$\pm$\halphaabsorbminimumwavestdgrism\,\AA}).
In contrast, the distribution of absorption minima we measure from the bootstrapped X-shooter spectra has a bimodal shape and is substantially broad, stretching over $\sim$ 15,650--16,050\,\AA. After removing a small number of outlying measurements close to 15,300\,\AA, we constrain the absorption minimum to be \mbox{\halphaabsorbminimumvelvlt$\pm$\halphaabsorbminimumvelstdvlt\,km\,s$^{-1}$} (\mbox{\halphaabsorbminimumwavevlt$\pm$\halphaabsorbminimumwavestdvlt\,\AA}). 

We next perform a data-driven simulation of the WFC3 grism spectrum to determine the statistical significance of finding the absorption feature we identify. 
As a first step, we smooth the grism spectrum using a $\sigma = 2000$\,km\,s$^{-1}$ Gaussian kernel, and calculate the residuals of the data from the smoothed spectrum in the wavelength range 14,000--16,100\,\AA. 
Each simulated spectrum is created by replacing the flux at each wavelength in the grism spectrum 
with a randomly drawn value from the distribution of residuals.

With 10,000 simulated spectra, we compute a test statistic that measures the strength of the absorption relative to the continuum.
To estimate the continuum level, we calculate ${\rm median}(f_{1.45-1.55})$, the median flux in the wavelength range 14,500--15,500\,\AA.
We next calculate ${\rm median}(f^{{\rm H}{\alpha}}_{\rm \,absorp})$, the median flux within $\pm150$\,\AA\ of the absorption minimum which corresponds to the 2--3$\sigma$ width of SN~1987A-like SN~II H$\alpha$ absorption features (see Figure~\ref{fig:grismmodel}). The difference, 
\begin{equation}
\Delta_{\rm absorp} = {\rm median}(f^{{\rm H}{\alpha}}_{\rm \,absorp}) - {\rm median}(f_{1.45-1.55}),
\end{equation}
is used as the test statistic to compute a $p$ value.
For the grism spectrum, we measure \mbox{ $\Delta_{\rm  absorp}^{\rm grism} = (-2.5\pm0.9) \times 10^{-20}$\,erg\,s$^{-1}$\,cm$^{-2}$\,\AA$^{-1}$}.
We compute the probability of finding a lesser value $\Delta_{\rm absorp}^{\rm grism}$ by random chance using the spectra we simulate.  We compute $p$ = \pvalueabsorptiongrism, which provides statistically significant evidence against the hypothesis that the apparent H$\alpha$ absorption feature in the grism spectrum is a random artifact.
 
Given the width of the H$\alpha$ absorption feature and $p$ = \pvalueabsorptiongrism\ significance, we would not expect any other similarly strong absorption feature, and none exists in the range 14,000--16,100\,\AA.
If we instead extend the wavelength range used in the analysis blueward to 11,500--15,500\,\AA, then 
the $p$-value increases to $\sim0.02$. 
Simulating grism spectra by repeatedly randomly drawing from residuals does not model any covariance in the random noise, although we do not expect a strong covariance.

In the case of the X-shooter spectra, the wavelength of the absorption minimum is poorly constrained, so it is not possible to apply the same statistical test. 
For the purpose of completeness, however, we calculate the $p$ value for an absorption feature located at  \halphaabsorbminimumwavevlt\,\AA, the median of the absorption minima measured from the bootstrapped spectra.
The continuum is measured as the median of the spectrum regions 15,300--15,600\,\AA\ and 16,500--17,000\,\AA. 
We calculate $p$ = \pvalueabsorptionvlt\ but note that \halphaabsorbminimumwavevlt\,\AA\ does not coincide with either peak of the bimodal distribution of minima measured from the bootstrapped spectra.

\subsection{H$\alpha$ Expansion Velocity Evolution and Comparison to SN~1987A-like SNe}
In Figure~\ref{fig:halphaminimum}, we plot the constraints on the H$\alpha$ expansion velocity, and compare the measurements against the expansion velocities of SN 1987A, SN 2006V, and SN 2006au measured by \citet{taddiastritzingersollerman12}.
The grism measurement at \grismphase\ favors an H$\alpha$ velocity comparable to that of the blue (see Figure~\ref{fig:ColorCurveComparison}) SN~1987A-like SN 2006V.  The approximate X-shooter constraint on the expansion velocity at \xshooterphase\ is consistent with the evolution of the H$\alpha$ expansion velocity of SN~1987A-like events.

\subsection{Models of the H$\alpha$ Emission and Absorption Profiles}
The H$\alpha$ emission from SNe~IIn generally exhibits a Lorentzian profile that arises from Thompson scattering of photons off of free electrons \citep{chugai01,smith10b}, while Doppler broadening of H$\alpha$ emission from SN~1987A-like SNe instead produces  approximately Gaussian profiles.
Since the line shape contains information about the SN spectroscopic type,
we examine which functional form better fits the grism and X-shooter spectra.
As shown in Figures~\ref{fig:grismmodel} and \ref{fig:vltmodel}, we also model the grism spectrum with a P-Cygni profile including an absorption feature. 

The model fits provide strong evidence favoring a Gaussian profile over a Lorentzian profile, and 
very strong evidence for a P-Cygni model for the WFC3 grism spectrum.
In Table~\ref{tab:aic}, we list the differences in the Akaike information criterion (AIC; \citealt{aka74}) to interpret the differences in the $\chi^{2}$ statistics. A change of 2 in the AIC provides evidence against the model having a greater AIC value, while a difference of 6 constitutes strong evidence (e.g., \citealt{kas95}; \citealt{muk98}). 
The AIC penalizes models having a greater number of parameters. 
A Gaussian model for the absorption feature in the grism spectrum yields a $\sim 4300$\,km\,s$^{-1}$ FWHM. 

\subsection{Measurements of the H$\alpha$ Line Profile and Flux}
In Figure~\ref{fig:halphastrength}, we compare the total H$\alpha$ luminosity of SN Refsdal at \grismphase\ and \xshooterphase\ with the H$\alpha$ luminosity of SN~1987A-like SNe, as well as SN 2005cp. 
The strength and change in the H$\alpha$ emission are consistent with the characteristics of H$\alpha$ emission from SN~1987A-like SNe.
We adopt a magnification of $\mu = 15$ to estimate the absolute luminosity, and error bars correspond to a 50\% uncertainty in the magnification (see Table~\ref{tab:magnifications}).

\begin{deluxetable*}{lcccc}
\tablecaption{Akaike Information Criteria}
\tablecolumns{5}
\tablehead{&\multicolumn{2}{c}{{\it HST} WFC3 Grism}&\multicolumn{2}{c}{VLT X-shooter}\\Model&$\Delta$AIC&$\chi^2$&$\Delta$AIC&$\chi^2$}
\startdata
Lorentzian & \AICLorentziangrism & \chisqLorentziangrism~($N_{\rm param} = $\numparLorentziangrism) &  \AICLorentzianvlt & \chisqLorentzianvlt~($N_{\rm param} = $\numparLorentzianvlt)\\
Gaussian & \AICGaussiangrism    & \chisqGaussiangrism~($N_{\rm param} = $\numparGaussiangrism) & \AICGaussianvlt & \chisqGaussianvlt~($N_{\rm param} = $\numparGaussianvlt)  \\
P Cygni & \AICPCygnigrism          & \chisqPCygnigrism~($N_{\rm param} = $\numparPCygnigrism) &  ...  & ...
\enddata
\tablecomments{Increments in the AIC for models of the H$\alpha$ emission and absorption. A difference greater than 6 is considered strong positive evidence against the model with the higher value.  We do not calculate a P-Cygni model for the VLT X-shooter spectrum, because we are only able to constrain the value of the minimum approximately. A Gaussian profile is favored over a Lorentzian profile for both the WFC3 grism and the X-shooter grism data, and a P-Cygni absorption feature is very strongly favored for the WFC3 grism spectrum.  }
\label{tab:aic}
\end{deluxetable*}

\begin{deluxetable}{lccc}
\tablecaption{Predicted Magnifications}
\tablecolumns{4}
\tablehead{\colhead{Model}&\colhead{$\mu_{\rm S1}$}&\colhead{$\mu_{\rm
S2}$}&\colhead{$\mu_{\rm S3}$}} 
\startdata
\citet{kellyrodneytreu15} & $\sim 10$ & $\sim 10$ & $\sim 10$ \\ 
\citet{oguri15} & 15.30 & 17.66 & 18.29 \\ 
\citet{sharonjohnson15} & 18.5$^{+6.4}_{-4.5}$ & 14.4$^{+7.5}_{-5.5}$ & 20.5$^{+19.1}_{-3.9}$ \\
\citet{grillokarmansuyu15} (G12F) & 16.0$^{+1.4}_{-5.7}$ & 14.3$^{+4.5}_{-6.4}$ & 15.2$^{+4.0}_{-4.9}$ \\ 
\citet{jauzacrichardlimousin15} & 22.4$\pm$2.0 & 18.9$\pm$2.3 & 19.7$\pm$1.7  
\enddata
\tablecomments{ Magnifications of images S1--S3 (for which we have spectra)
predicted by models of the combined gravitational potential of the early-type
galaxy and the MACS1149 galaxy cluster lenses. }
\label{tab:magnifications}
\end{deluxetable}

\begin{deluxetable*}{lcccccc}
\tablecaption{Nearby Supernovae with SN~1987A-like Light Curves}
\tablecolumns{7}
\tablehead{\colhead{SN} & \colhead{Host} & \colhead{Milky Way} & $D$ & $R$- or
$r$-band &\colhead{Spectroscopy} & \colhead{Photometry} \\
\colhead{} &  $E(B-V)$ (mag) & $E(B-V)$  (mag) & (Mpc) & Max. (MJD) & Dataset &
Dataset
}
\startdata
SN 1987A & 0.13 (1) & 0.06 (2) & 0.50$\pm$0.005 & 46933.10$\pm$1.0 & 3 & 4  \\
SN 1998A & $\sim 0$ (5) & 0.12 & 33$\pm$10 &  50885.10$\pm$3.9 & 5 & 6 \\
SN 2005cp & 0.02 (6) & 0.03 & 120$\pm$9.9  & 53581 & 7,8 & 7 \\
SN 2006V & $\sim 0$ (7) & 0.029 & 72.7$\pm$5 & 53824.23 & 9 & 9 \\
SN 2006au & 0.141 (8) & 0.172 & 46.2$\pm$3.2 & 53866.25   & 10 & 10 \\
SN 2009E & 0.02 (8) & 0.02 & 29.97$\pm$2.10 (8) & 54927.8$\pm$2.8 &  10 & 10
\enddata
\tablecomments{Publications containing data used for comparison. Many of the
spectra were retrieved from WISEREP\footnote{http://wiserep.weizmann.ac.il/}
\citep{yarongalyam12}.  \citealt{weltyxuewong12} (1);
\citealt{staveleysmithkimcalabretta03} (2); \citealt{hamuysuntzeff90} (3);
\citealt{phillipsheathcotehamuy88}\footnote{http://www.physics.unlv.edu/\~jeffery/astro/sne/spectra/d1980/sn1987a/old/}
(4); \citealt{phillipshamuyheathcote90}\footnote{ftp://ftp.noao.edu/sn1987a}
(5); \citealt{pastorellobaronbranch05} (6); 
\citealt{kiewegalyamarcavi12} (7);  
\citealt{bsnipi} (8);
\citealt{taddiastritzingersollerman12} (9); 
\citealt{pastorellopumonavasardyan12} (10). Distances taken from NASA/IPAC
Extragalactic Database\footnote{https://ned.ipac.caltech.edu/} except if another
citation is provided.  Milky Way extinction from
\citet{schlaflyfinkbeinerSFD11}, except for the case of SN 1987A in the LMC.   }
\label{tab:comparisondata}
\end{deluxetable*}

\subsubsection{Comparison with Spectra of Other Supernovae}
In Figures~\ref{fig:grismsncomp} and \ref{fig:vltsncomp}, we compare the grism 
and X-shooter data with spectra of SN~1987A-like SNe as well as 
SN 2005cp at a similar phase. 
For each comparison spectrum, we also calculate the $\chi^2$ agreement with the spectrum of SN Refsdal in the wavelength range 15,000--17,000\,\AA.

To perform a comparison, we scale the spectrum of the low-redshift SN so that its synthetic $F160W$ flux matches the average flux of SN Refsdal when it was observed.
While we have already subtracted the galaxy contribution from the grism spectrum at an earlier step,
we need to model the underlying galaxy light in the X-shooter spectrum.
We use the spectrum of an Sc galaxy redshifted to $z=1.49$ (spiral host galaxy) as a model the galaxy light, and 
vary its normalization to find the best match to the data.
Using the spectrum of an S0 galaxy at $z=0.54$ (early-type galaxy lens) yields almost identical results.
All of the low-redshift comparison spectra are corrected for the Milky-Way and host-galaxy extinction values listed in Table~\ref{tab:comparisondata}.

\subsection{Constraint on the Supernova Ejecta Mass}
Following \citet{taddiastritzingersollerman12}, we scale the parameters of the \citet{blinnikovlundqvistbartunov00} model of SN 1987A to estimate the ejecta mass of SN Refsdal.
To scale the model, we use the relation $t_d \approx (\kappa M_{\rm ej}/v)^{1/2}$ from  \citet{arnett79},
where $t_d$ is the diffusion time, $\kappa$ is the mean opacity, and $v$ corresponds to the expansion velocity. 
We lack direct constraints on the explosion date and the time of the bolometric peak, 
but SN~1987A-like SNe exhibit a small dispersion in their rise times; see Table~5 of \citet{pastorellopumonavasardyan12}.
These assumptions and a scaling according to the H$\alpha$ expansion velocity yield an estimate for the ejecta mass of \ejectamass.

\begin{deluxetable}{lcccc}
\tablecaption{Best-Fitting Template Spectra in Superfit Library}
\tablecolumns{5}
\tablehead{\colhead{Supernova}&\colhead{Type}&\colhead{Phase}&\colhead{$A_V$ (mag)}&\colhead{$S$}}
\startdata
SN 1998A & II & -40 & 1.1 & 44.62 \\
SN 2005cs & II & +2 & 1.9 & 48.98 \\
SN 2005cs & II & +14 & 0.9 & 49.05 \\
SN 2004et & II & +47 & -0.9 & 49.20 \\
SN 1999em & II & +20 & -0.2 & 49.45 \\
SN 1986I & II & +83 & 1.1 & 49.70 \\
SN 2004et & II & +45 & -0.7 & 49.79 \\
SN 2005cs & II & +1 & 1.9 & 49.90 \\
SN 2005cs & II & +6 & 0.9 & 49.98 \\
SN 2005cs & II & +11 & 0.5 & 50.45 \\
SN 1999em & II & +9 & -0.2 & 50.48 \\
SN 1999em & II & +15 & -0.4 & 50.50 \\
SN 1999em & II & +4 & 1.7 & 50.79 \\
theory99em & II & +25 & -0.7 & 50.85 \\
SN 1993W & II & +21 & -0.1 & 50.85 
\enddata
\tablecomments{The fifteen best-matching {\tt Superfit}
\citep{howellsullivanperrett05} templates are SNe~II showing P-Cygni profiles, which are
characterized by the presence of H features. $S=\Sigma (F_{i}-T_{i} \times
10^{-A_{\lambda}/2.5}) / \sigma_i^2$, where $F_i$ is the $F_{\lambda}$ flux of
the SN measured in each resolution element, $\sigma_i$ is the uncertainty of the
measured flux, $T_i$ is the flux of template spectrum, and $A_{\lambda}$ is the
wavelength-dependent extinction for an $R_V=3.1$ law. The normalization of the
template spectrum and the extinction $A_V$ are the fitting parameters.  We fix
the redshift to be that of the host galaxy ($z=1.49$).  }
\label{tab:superfitmatches}
\end{deluxetable}

\subsection{{\tt Superfit} Analysis of Grism Spectrum}
To identify the spectroscopic classification of a SN, the {\tt Superfit} \citep{howellsullivanperrett05} tool computes the $\chi^2$ agreement between an input spectrum and a set of template SN spectra\footnote{https://github.com/dahowell/superfit} reddened (or dereddened) by a range of $A_V$ values. If the redshift of the SN is uncertain, template spectra can also be shifted across the possible redshifts of the SN. To calculate the $\chi^2$ statistic, {\tt Superfit} can use the uncertainties of the values in the spectrum. 

We apply {\tt Superfit} to find the SN that best matches the grism spectrum of SN Refsdal. Given the similarity of the light curve of SN Refsdal to those of SN~1987A-like SNe, we add spectra of SN 1998A, SN 2005cp, SN 2006V, SN 2006au, and SN 2009E to the set of {\tt Superfit} template spectra. Since the position of the lensed source is near the tip of a spiral arm of a lensed galaxy, we fix the SN redshift to that of the host ($z=1.49$) to investigate whether we can obtain a satisfactory fit. We allow the extinction $A_V$ to vary between $-$2 and $+$2 mag, and we do not include any contribution from the host galaxy since that has already been subtracted. Our input spectrum is not binned and includes the flux measurements taken in both telescope orientations, as well as the uncertainties in the measured fluxes. We adopt five iterations of $5\sigma$ outlier rejection, which is less aggressive than the default 2.7$\sigma$ rejection.

Table~\ref{tab:superfitmatches} lists the fifteen best-fitting template matches to the grism spectrum in the wavelength range 14,000--16,750\,\AA. These include only SNe~II having H$\alpha$ P-Cygni profiles.  If we extend the wavelength range to 11,000--16,750\,\AA, then a spectrum of SN 2005cp taken $\sim 30$--50 days after maximum light provides the fifth-best match. All other matches are SNe~II having P-Cygni profiles.  While the template spectrum of SN 2005cp was taken $\sim 30$--50 days after maximum light, the grism spectra of SN Refsdal were acquired at \grismphase, and the H$\alpha$ profile of SN 2005cp broadened significantly after maximum light as CSM interaction increased (see Figure~9 of \citealt{kiewegalyamarcavi12}). 
As shown in Figure~\ref{fig:halphastrength}, the H$\alpha$ evolution of SN Refsdal does not appear to be the same as that of SN 2005cp.

\section{The SN Host Galaxy and Environment}
\label{sec:snenvironment}

Since the progenitor of SN 1987A was identified as a  blue supergiant star \citep{gilmozzicassatellaclavel87,sonnebornaltnerkirshner87}, the precursors of SN~1987A-like SNe are believed to be similar --- compact, short-lived, massive stars. 
The properties of the emitting gas near the explosion site of SN Refsdal measured from strong nebular lines should be similar to the properties of the gas that formed the massive progenitor of SN Refsdal.
To measure the host-galaxy narrow-line emission, we reduce the spectra in ``stare mode'' where we do not subtract the off-target spectra.  
This allows us to avoid subtracting any narrow-line emission from sources in the off-target position which can have bright emission lines.
We list the narrow-line measurements in Tables~\ref{tab:emissionlinesmosfire} and \ref{tab:emissionlinesxshooter} and the inferred extinction and properties of the ionized gas in Table~\ref{tab:gasproperties}. 

As Figure~\ref{fig:VLTexpsetup} shows, we detect strong nebular emission   
both near the SN explosion site and from the nuclear region of the host galaxy. 
The emission from the nuclear region has a significantly broader profile, which we attribute to the rotational motion of the gas. From the best-fit line centers,
we measure redshifts of $1.48831\pm0.00007$ from the spectrum extracted around the SN position and $1.48837\pm0.00007$ for the nuclear region,  
where the uncertainties include contributions from both line fitting and the wavelength solution.

To constrain the source of the ionizing radiation and properties of the emitting gas, we construct a Baldwin, Phillips, \& Terlevich (BPT; \citealt{baldwin81}) diagram, plotted in Figure~\ref{fig:BPT}.
The intensity ratios [\ionpat{O}{iii}]$\lambda$5007/H$\beta$ and [\ionpat{N}{ii}]$\lambda$6584/H$\alpha$ of the strong emission lines from near the SN site and the nuclear region are 
consistent with the ratios expected for gas ionized by radiation from massive stars, and the positions on the BPT diagram coincide with the Sloan Digital Sky Survey (SDSS; \citealt{thomassteelemaraston13}) star-forming galaxy population.

\begin{figure*}
\centering
\subfigure{\includegraphics[angle=0,width=6.5in]{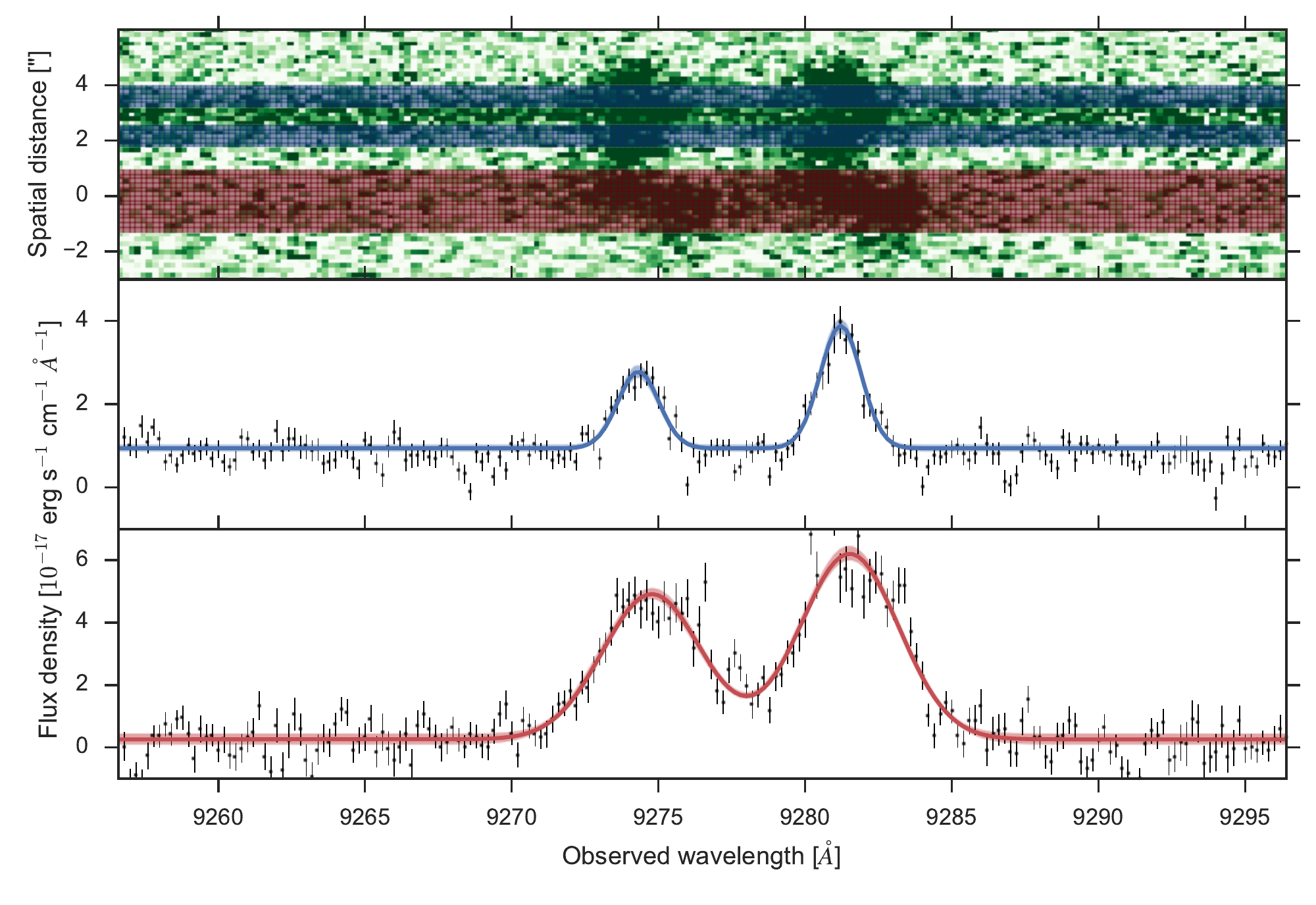}}
\caption{Example of nebular emission line detection in VLT/X-shooter spectra. The top panel is the two-dimensional
spectrum from OB2 (see Figure~\ref{fig:slits}) in the region surrounding
[\ionpat{O}{ii}] $\lambda\lambda$3726, 3729. The ordinate corresponds to the spatial direction and the abscissa is the spectral direction. The blue-colored regions are the extraction apertures used for the
position of the SN images S1 and S2 (above and below, respectively). 
The trace visible between the two SN positions is light from the early-type galaxy lens. 
The red-colored region is the aperture used to extract the spectrum of the host nucleus. 
The [\ionpat{O}{ii}] $\lambda\lambda$3726, 3729 doublet is
cleanly resolved in the X-shooter spectrum, and significant structure is
visible in both the spatial and dispersion directions. The middle panel shows the fit to
the [\ionpat{O}{ii}] $\lambda\lambda$3726, 3729 emission lines at the SN positions, while 
the bottom panel plots the same for the host nucleus.}
\label{fig:VLTexpsetup}
\end{figure*}

\begin{deluxetable*}{lc}
\tablecaption{Keck-I MOSFIRE Host-Galaxy Emission-Line Measurements}
\tablecolumns{2}
\tablehead{\colhead{Line}&\colhead{Flux}\\
\colhead{}&\colhead{($10^{-17}$\,erg\,s$^{-1}$\,cm$^{-2}$)}
}
\startdata
\nii\ $\lambda$6549 & $0.197 \pm 0.0598$  \\
H$\alpha$                                &  $5.87 \pm 0.132$ \\
\nii\ $\lambda$6584 &  $0.592 \pm 0.179$ \\
\sii\ $\lambda$6717 & $1.34 \pm 0.141$ \\
\sii\ $\lambda$6731 & $0.854 \pm 0.147$ 
\enddata
\tablecomments{Measurements of emission lines and uncertainties. Fluxes are not
corrected
for extinction or magnification.}
\label{tab:emissionlinesmosfire}
\end{deluxetable*}

\begin{deluxetable*}{cccccc}
\tablecaption{VLT X-shooter Host-Galaxy Emission-Line Measurements}
\tablecolumns{7}
\tablehead{
\colhead{Source}&\colhead{Line}&\colhead{Line Flux}&\colhead{FWHM} \\
\colhead{}&\colhead{}&\colhead{($10^{-17}$\,erg\,s$^{-1}$\,cm$^{-2}$)}
&\colhead{(km\,s$^{-1}$)}
}
\startdata
SN & \oii\ $\lambda$3726  & 0.47 $\pm$ 0.08 & 43 $\pm$ 18 \\
SN & \oii\ $\lambda$3729  & 0.63 $\pm$ 0.01 &  -  \\
SN & \hb & 0.2 $\pm$ 0.1 & 64$\pm$ 16 \\
SN & \oiii\ $\lambda$4959  & 0.25 $\pm$ 0.05 & 60 $\pm$ 14 \\
SN & \oiii\ $\lambda$5007  & 0.7 $\pm$ 0.1 &  -  \\
SN &  \ha   & 1.5 $\pm$ 0.3 & 66 $\pm$ 11 \\
SN &  \nii\ $\lambda$6584   &  $\leq$0.24  & 62 $\pm$ 13 \\
\hline         
Host nucleus & \oii\ $\lambda$3726  & 3.1 $\pm$ 0.2 & 185 $\pm$ 18 \\
Host nucleus & \oii\ $\lambda$3729  & 3.6 $\pm$ 0.2 &  -  \\
Host nucleus & \hb & 4.4 $\pm$ 0.6 & 168 $\pm$ 7 \\
Host nucleus & \oiii\ $\lambda$4959  & 0.9 $\pm$ 0.2 & 177 $\pm$ 20 \\
Host nucleus & \oiii\ $\lambda$5007  & 1.0 $\pm$ 0.2 &  -  \\
Host nucleus &  \ha   & 14.0 $\pm$ 0.2 & 203 $\pm$ 30 \\
Host nucleus &  \nii\ $\lambda$6584   & 4 $\pm$ 1 & 187 $\pm$ 18 
\enddata
\tablecomments{Measurements of emission lines and uncertainties. Fluxes are not
corrected for extinction or magnification. FWHM line widths are corrected for the instrumental
resolution. For the instrumental resolution, the nominal value for the 0\farcs6
slit is used since the FWHM of the PSF is comparable for all observations,
so the resolution is set by the seeing rather than the slit width.}
\label{tab:emissionlinesxshooter}
\end{deluxetable*}

\begin{deluxetable*}{cccc}
\tabletypesize{\scriptsize}
\tablecaption{Host-Galaxy Measurements
\label{tab:gasproperties}}
\tablecolumns{4} 
\tablehead{
	\colhead {Measurement} & 
	\colhead {Nuclear Region} &
	\colhead {SN Site} &
	\colhead {Instrument} 
 } 
\startdata
12 + log(O/H) (PP04 N2)  &  ... & $8.3 \pm 0.1$\,dex & MOSFIRE  \\
12 + log(O/H) (PP04 N2)  &  $8.6 \pm 0.1$ dex & 8.4\,dex (3$\sigma$ upper limit) & X-shooter \\
$E(B-V)$  & $0.1 \pm 0.2$ mag & $0.8 \pm 0.4$ mag & X-shooter \\
$A_{{\rm H}\alpha}$  &  $0.2 \pm 0.3$ mag & $1.5 \pm 0.9$\,mag & X-shooter \\
$A_{V}$  &  0.3$ \pm $0.4 mag & 2.0$ \pm $1.0\,mag  & X-shooter 
\enddata
\tablecomments{Properties of the ionized gas near the SN explosion site and in the host-galaxy nuclear region, as well as reddening along the line of sight from analysis of strong nebular emission lines. }
\end{deluxetable*}

We next estimate the reddening along the line of sight to the emitting gas from the measured Balmer decrement.
Making the assumption of Case B recombination, we apply the prescription from \citet{dominguezsianahenry13} and adopt the $R_{V} = 2.51$ \citet{reddykriekshapley15} extinction curve inferred from MOSFIRE Deep Evolution Field (MOSDEF) spectroscopy of galaxies at $z \approx 1.4$--2.6.
We estimate a color excess of $E(B-V) = 0.1\pm0.2$ mag ($A_{V} = 0.3 \pm 0.4$\,mag) for the nuclear region and $E(B-V) = 0.8\pm0.4$ mag ($A_{V} = 2.0 \pm 1.0$\,mag) near the SN site.
In Table~\ref{tab:gasproperties}, we also list the extinction expected for H$\alpha$. 

SN Refsdal occurred at an offset from its host galaxy's nucleus of $\sim7$\,kpc, and \citet{yuankobayashikewley15}
used OSIRIS \citep{larkinbarczyskrabbe06} integral-field-unit observations to constrain the oxygen abundance from $\sim5$--7\,kpc to be 12 + log(O/H)$_{\rm
PP04N2}$ $\leq$ 8.11\,dex by comparing an upper limit on the [\ionpat{N}{ii}] emission to the
measured H$\alpha$ flux using the \citet{pettini04} abundance diagnostic. 
As in \citet{yuankobayashikewley15}, we use the \citet{pettini04} N2 metallicity indicator based on the ratio [\ionpat{N}{ii}] $\lambda$6584/H$\alpha$. 
We find that the nuclear region has an oxygen abundance of \mbox{12 + log(O/H) = $8.6 \pm 0.1$ dex}. From an upper limit on the [\ionpat{N}{ii}] flux, we compute a 3$\sigma$ upper limit of 8.4\,dex near the SN site.    
The MOSFIRE spectrum yields a 3$\sigma$ detection of [\ionpat{N}{ii}] and an estimate for the oxygen abundance of 12~+~log(O/H)~$= 8.3 \pm 0.1$ dex.
The presence of strong night-sky lines close to both \nii\ emission lines limit the sensitivity of the spectra to the metallicity.
Since the ratio of nitrogen to oxygen and the ionization parameter substantially affect the N2 metallicity diagnostic,
our estimates may have an accuracy of only $\sim0.2$\,dex. 
The relatively large uncertainty in our measurement of H$\beta$ implies that the $R_{23}$ diagnostic \citep{pagel79} is not able to provide a useful constraint on the oxygen abundance.

\begin{figure*}
\centering
\subfigure{\includegraphics[angle=0,width=5.25in]{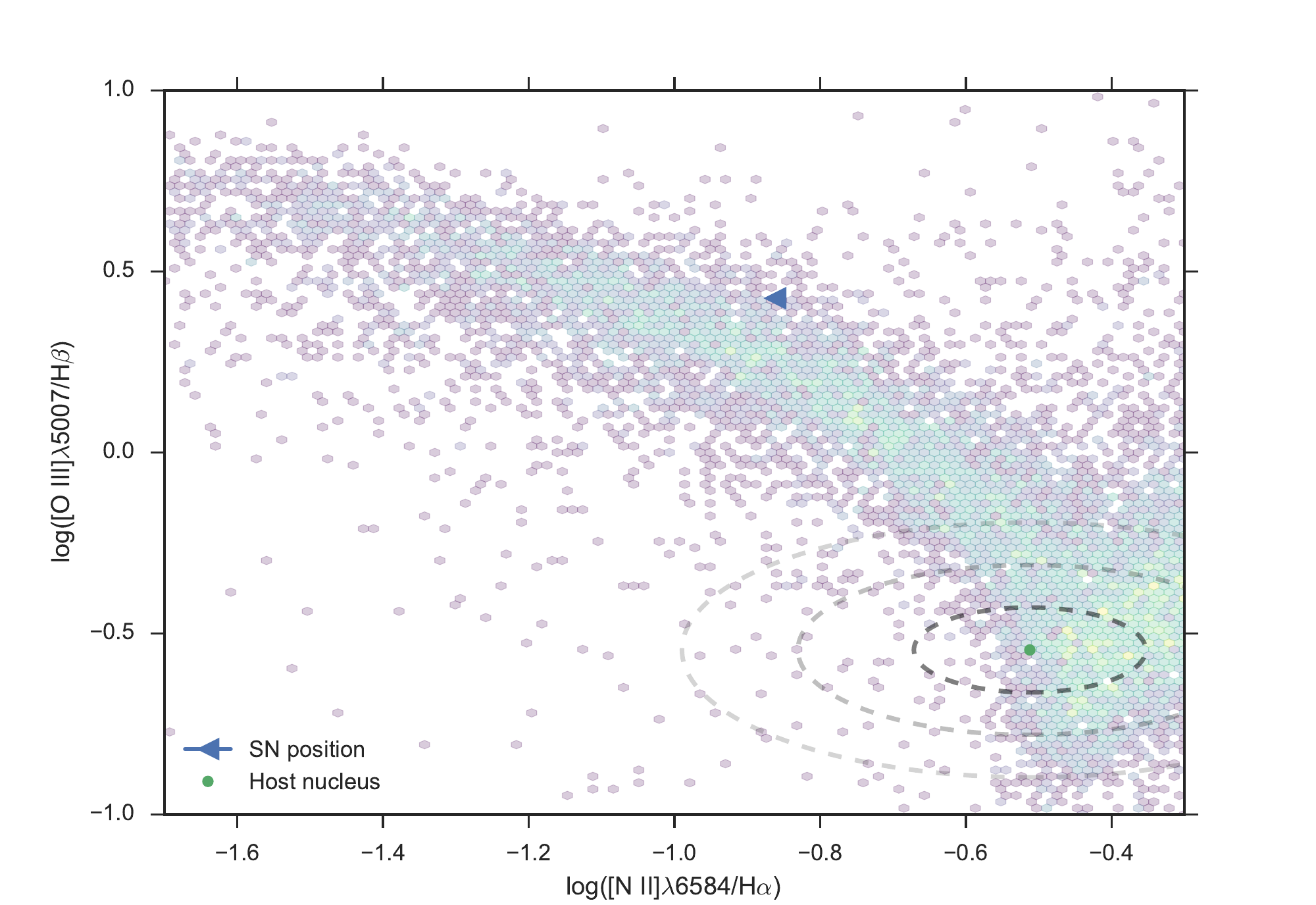}}
\caption{BPT diagram \citep{baldwin81} for the emission lines measured from two host-galaxy extractions. The green dot shows the ratios for the host nuclear region, where the ellipses correspond to 1$\sigma$, 2$\sigma$, and 3$\sigma$ confidence contours. The blue triangle marks the ratios of emission lines originating from near the SN position, where only an upper limit for \nii~flux could be estimated. This turns into an upper limit on the log(\nii/\ha) ratio. Underplotted is the same ratio for all the $z \leq 0.1$ objects in the SDSS DR12 \citep{thomassteelemaraston13}.}
\label{fig:BPT}
\end{figure*}

\citet{karmangrillobalestra16} have used the Multi Unit Spectroscopic Explorer (MUSE) on the VLT to
take optical spectra of the three separate images of the host-galaxy environment of SN Refsdal. Within a 0\farc6-radius aperture surrounding the explosion site, they report a high ratio of \oii\ to \mgii\ of 10--20, consistent with low metallicity and a high ionization parameter.
From the X-shooter spectra, we measure $3\sigma$ limits on the flux of \mgii\ $\lambda2796$ to be $\leq 4.7 \times 10^{-18}$  erg
s$^{-1}$ cm$^{-2}$ and of \mgii\ $\lambda2803$ to be $\leq 6.3 \times 10^{-18}$  erg
s$^{-1}$ cm$^{-2}$, which are consistent with the flux levels measured by
\citet{karmangrillobalestra16}.

\subsection{Comparison of Host-Galaxy Environments}

The low oxygen abundance inferred for the explosion environment of SN Refsdal 
is consistent with a preference for low metallicity among low-redshift SN~1987A-like SNe.
\citet{taddiasollermanrazza13} found an average value of 12 + log(O/H) = 8.36 $\pm 0.05$\,dex (PP04 N2) for the local environments of SN~1987A-like SNe. 
They found that these abundances were more metal-poor than those of SN~IIP 
environments (within $\sim3$\,kpc) measured by \citet{anderson10}.
Additional circumstantial evidence for low-metallicity environments include a preference for
low-mass dwarf host galaxies, or the periphery of late-type spirals \citep{pastorellopumonavasardyan12}.
By contrast, normal SNe~II trace the light $g$-band light of their host galaxies \citep{kel08} 
and show no preference for metal-poor galaxies or the peripheries of their host galaxies \citep{kelkir12}.

\section{Summary}
\label{sec:summary}

We have used {\it HST} WFC3 grism and VLT X-shooter NIR
spectra and images to classify SN Refsdal, the first 
example of a resolved, strongly lensed SN.
Its slowly rising broadband {\it HST} light curve can be matched by those of SN~1987A-like SNe, a
peculiar class of SNe~II that may account for
$\sim 1.5$--3\% of nearby SNe \citep{smartteldridge09,kleiserpoznanskikasen11,pastorellopumonavasardyan12}.
The only other SNe having similar light curves and colors are SNe~IIn, which erupt into dense CSM and whose light curves are heterogeneous.
Detection of strong and wide H$\alpha$ P-Cygni absorption in the grism spectrum, however, identifies SN Refsdal as a SN~1987A-like SN, and excludes the possibility that it is a SN~IIn with continuum emission arising primarily from circumstellar interaction.

The specific properties of the H$\alpha$ emission and absorption features are 
also consistent with measurements of low-redshift SN~1987A-like SNe.
The H$\alpha$ expansion velocity we measure from the grism spectrum agrees with
the velocities of SN~1987A-like SNe at similar pre-maximum phases. 
Similarly, the strength and evolution of the H$\alpha$ luminosity are consistent with those of 
SN~1987A-like events, and may not have been as strong or evolved the same way as was observed for the Type IIn SN 2005cp. 
Finally, the H$\alpha$ emission-line profile is Gaussian instead of Lorentzian.

SN 1987A was the nearest detected SN of the modern era and is the
best-studied event.
Its progenitor star (Sk $-69^{\circ}$202) in the LMC 
was a compact  blue supergiant (B3~I;
\citealt{gilmozzicassatellaclavel87,sonnebornaltnerkirshner87})
inferred to have a mass close to 20\,M$_{\odot}$ (see
\citealt{arnettbahcallkirshner89} for a review).
The size of the progenitor ($\lesssim100$\,R$_{\odot}$) was significantly smaller
than those
of average red supergiants (500--1000\,R$_{\odot}$), and accounts for the slowly
rising light curve.
\citet{pastorellopumonavasardyan12} suggest that all known SN~1987A-like SNe are
consistent with explosions of blue supergiant progenitors with radii between 35
and 90\,R$_{\odot}$ and masses at explosion of $\sim20$\,M$_{\odot}$, 
which would be higher, on average, than those identified for SN~IIP progenitors
\citep{smartt09}.

For assumed values of the magnification, SN Refsdal had a high luminosity in comparison to well-studied low-redshift SN~1987A-like SNe \citep{pastorellobaronbranch05, kleiserpoznanskikasen11,
taddiastritzingersollerman12, pastorellopumonavasardyan12}.  We find that the $V-R$ color of SN Refsdal is similar to those of low-redshift SN~1987A-like SNe, but that it has a comparatively blue $B-V$ color, especially near maximum light. Scaling from a model of SN 1987A and using the fact that SN~1987A-like SNe exhibit a relatively small range of rise times, we estimate an ejecta mass of \ejectamass\ for the explosion.

We note that circumstellar interaction can range in intensity from very weak to very strong depending on the distribution, mass, and composition of any material surrounding the star.   In the case of PTF 11iqb, for example, the CSM was overtaken early by the expanding ejecta and early narrow lines disappeared \citep{smithmauerhancenko15}.  Despite the lack of narrow lines near maximum light in spectra of PTF 11iqb, the CSM likely caused the SN to have a bluer color and higher luminosity.   

Indeed, SN~1987A is surrounded by a ring thought to have been ejected approximately 10$^4$ yr before the explosion \citep{meaburnbryceholloway95,crottsheathcote00}. \citet{smithmauerhanprieto14} have found that the light curves and spectra of the Type IIn SN~2009ip and SN~2010mc could be explained by the faint SN~1987A-like explosions of blue supergiant progenitors into dense CSM. It is possible that a small or even modest amount of circumstellar interaction could contribute to the color and luminosity of SN Refsdal that we observe.
 
Ongoing late-time {\it HST} imaging of the four images of SN Refsdal (PI P. Kelly; GO-14199) will continue to improve our understanding of the SN.
As predicted \citep{kellyrodneytreu15,oguri15, sharonjohnson15,diegobroadhurstchen15,jauzacrichardlimousin15,treubrammerdiego15,grillokarmansuyu15,kawabataoguriishigaki15}, SN Refsdal has reappeared in a different image of its spiral host galaxy \citep{kellyreappear15}.  
While other faded SNe depart forever, we will have a second opportunity to study SN Refsdal and to learn about its early evolution.

\acknowledgements
We would like to thank Space Telescope Science Institute (STScI) director Matt
Mountain for making it possible to obtain the WFC3 grism spectra of SN Refsdal. 
Based on observations made with ESO Telescopes at the La Silla Paranal Observatory under program ID 295.D-5014.
We express our appreciation for the efforts of {\it HST} Program Coordinator Beth Periello
and Contact Scientist Norbert Pirzkal, as well as Claus Leitherer, Andrew Fox, Ken
Sembach, and Miranda Link for 
scheduling the {\it HST} grism observations. 
We thank Francesco Taddia for sharing spectra of SN 2006V and SN 2006au
with us, Ori Fox for helpful discussions about the Type IIn SN 2005cp, and Nathan Smith for useful comments about CSM interaction and SN~1987A-like SNe. 

Support for the preparation of the paper is from {\it HST} grants
GO-14041 and GO-14199. The GLASS program was supported by GO-13459, and the
FrontierSN photometric follow-up program has funding through GO-13386.  
A.Z. is supported by a Hubble Fellowship (HF2-51334.001-A) awarded by STScI, which is operated by
the Association of Universities for Research in Astronomy, Inc., for
NASA, under contract NAS 5-26555.  R.J.F. gratefully acknowledges support
from National Science Foundation (NSF) grant AST-1518052 and the Alfred P. Sloan Foundation.  A.V.F.'s
group at UC Berkeley has received generous financial assistance from
the Christopher R. Redlich Fund, the TABASGO Foundation, Gary \&
Cynthia Bengier, and NSF grant AST-1211916.   Supernova research at Rutgers
University (S.W.J.) is supported by NSF CAREER award AST-0847157, and NASA/Keck
JPL RSA 1508337 and 1520634.
C.M. is supported through NSF grant AST-1313484.
The DNRF provided funding for the Dark Cosmology Centre. 
J.M.S. is supported by an NSF Astronomy and Astrophysics Postdoctoral
Fellowship under award AST-1302771.

This research uses services or data provided by the NOAO Science Archive. NOAO
is operated by the Association of Universities for Research in Astronomy (AURA),
Inc., under a cooperative agreement with the NSF.
Some of the data presented herein were obtained at the W. M. Keck
Observatory, which is operated as a scientific partnership among the
California Institute of Technology, the University of California, and
NASA; the observatory was made possible by the generous financial
support of the W. M. Keck Foundation. We recognize the Hawaiian 
community for the opportunity to conduct these observations from the summit
of Mauna Kea.

\appendix

\section{Methods}
\label{sec:methods}

\subsection{(1) Phases of the MOSFIRE, WFC3, and X-shooter Spectra Relative to Time of Maximum Light}
 
The mean phase relative to maximum light (in the SN rest frame) of the Keck-I
MOSFIRE data is \mosfirephase,
of the {\it HST} WFC3 G141 grism spectra is \grismphase, and
of the VLT X-shooter spectra is \xshooterphase.
At maximum light, $F160W \approx 24.45$\,mag AB for image S2.

\subsection{(2) {\it HST} WFC3-IR G141 Grism Data}

\subsubsection{(a) Processing}
We processed the WFC3/IR FLT images obtained from the {\it HST} archive using the
software pipeline developed for the 3D-HST
project\footnote{https://github.com/gbrammer/threedhst}
\citep{brammervandokkumfranx12}.
This pipeline employs the {\tt DrizzlePac} package \citep{gonzaga12}\footnote{http://adsabs.harvard.edu/abs/201two-dimensionalrzp.book.....G} to
align the WFC3 images, and a custom software package to extract and model the grism spectra \citep{brammer12b, momcheva15}\footnote{http://adsabs.harvard.edu/abs/2012ApJ...758L..17B}\footnote{http://adsabs.harvard.edu/abs/2015arXiv151002106M}.

\subsubsection{(b) Modeling Contaminant Spectroscopic Traces}
In a grism spectral element, light passes first through a prism and then a
diffraction grating. 
The light from all objects in the field is dispersed in a common direction,
making  
efficient spectroscopy of a large number of objects possible. 
The Einstein cross, however, presents a challenge for grism spectroscopy,
because the SN images are embedded in light from both the spiral host galaxy and
the early-type galaxy lens.
In Figure~\ref{fig:grismorient}, we show a coaddition of all $F125W$ and
$F160W$ direct exposures of the SN images S1--S4, adjacent sources, and the
dimensions of the first-order and second-order grism traces.
The traces of nearby, bright cluster members create additional strong
contamination, and a nearby bright star with \starmag\ produces strong
diffraction spikes.

The roll angle of the spacecraft determines the dispersion axis of the WFC3 G141
grism. 
Acquiring spectra at more than one roll angle makes possible a more robust
extraction, because the trace at each roll angle is contaminated by different
sources. 
To identify the two combinations of roll angles and SN images that would
yield the least contamination, we simulated the expected grism spectra at all
roll angles available at the time of the observation, and predicted that 
image S2 observed at angles \one\ and \two\ would be least contaminated.

Even after configuring the observations to obtain the cleanest possible SN
spectra, 
light from other sources accounts for 85--90\% of the light
along the traces of image S2 of SN Refsdal. 
To recover the SN spectrum, 
we constructed and subtracted models of overlapping traces. We used the wide-band $F125W-F160W$ color and the $F160W$ flux to
model the continuum of the spiral host galaxy, the $z = 0.54$ elliptical
cluster-member lens, and the two adjacent red-sequence cluster members.
We used the $F160W$ flux to model the H$\alpha$ and \oiii\ narrow-line emission from the $z = 1.49$ spiral host galaxy.
Figure~\ref{fig:twod} shows the two-dimensional models for the contaminating
sources. 
Finally, we subtracted the residual background measured within a parallel adjacent
aperture.
Figure~\ref{fig:contamination} shows the successive removal of the modeled
sources
and the parallel background measurement for the grism spectrum taken in the
\one\ telescope orientation.
In Figure~\ref{fig:orientsoned}, we plot the extracted spectra of image S2 in
orients \one\ and \two, and, for comparison, that of SN 1987A at the same epoch.

\subsubsection{(c) Rejecting Outlying Flux Measurements}
We compute a smoothed spectrum at the wavelength of each measurement with a
weight proportional to a Gaussian density with $\sigma = 3000$\,km\,s$^{-1}$ in
the rest frame and inversely proportional to the square of the uncertainty 
of the flux measurement \citep{tonrydavis79}. 
The narrow lines have already been removed during an earlier processing step.

\subsubsection{(d) Comparison Between Direct Imaging Magnitudes and Grism Spectra}
\label{sec:specphot}
Given the fact that the SN flux was only a small fraction of the 
contaminating flux along the trace, 
a useful consistency check is to assess
how well synthetic magnitudes calculated from the grism SN spectrum agree with
direct photometry of the SN.
We compute synthetic AB magnitudes using the $F125W$ and $F160W$ filter
functions,
and calculate a synthetic $F125W-F160W$ color.

The grism spectra were taken from \grism, but all of the 202.9\,s post-imaging
exposures 
were taken with the $F160W$ wide-band filter from 23--28 December, while 
all post-imaging exposures from 30 December 2014 through 4 January 2015 
was taken with the $F125W$ wide-band filter.  
To estimate the average color of SN Refsdal during the period of grism
observations, we 
have coadded post-imaging exposures acquired on \grismphot, when coverage spans
approximately the same
epochs. These dates bracket the midpoint of the grism observations on \grismmid.

We measure direct magnitudes of $F125W = 25.01 \pm 0.05$\,mag AB and $F160W = 25.06\pm0.05$\,mag, or a color of $F125W-F160W = -0.05 \pm 0.07$\,mag AB. 
Then we calculate a synthetic color of $F125W-F160W = 0.14$\,mag AB. 

\subsection{(3) VLT X-shooter Spectroscopy}
The VLT data were reduced using the ESO/X-shooter pipeline v2.5.2
\citep{modiglianigoldoniroyer10}, where the Reflex interface
\citep{freudlingromaniellobramich13} is used
to manage the pipeline. The two-dimensional spectra have been rectified on a
grid with 0.2\,\AA\,pix$^{-1}$ in the ultraviolet blue (UVB) and VIS and 0.6\,\AA\,pix$^{-1}$ in the NIR
arm, slightly oversampling the spectra given the nominal X-shooter resolving power. 
The spectra are flux calibrated using an observation
of a spectrophotometric standard \citep{vernetkerbermainieri10,
hamuysuntzeffheathcote94} from which we measure a
response function.

The cluster field in which data were taken contains light from the $z=1.49$ spiral host galaxy 
and cluster galaxies in the
off-region locations observed during the nodding sequence.
The continua of these sources are relatively featureless near the H$\alpha$ emission from the SN, so we do not expect subtraction of the emission from these contaminants to affect the VLT SN spectrum.
While there is no evidence of subtraction of significant narrow-line nebular emission 
for OB1 and OB2, line emission in the ``off region'' used for OB3 contains strong emission, and we do not use the spectra acquired during this OB when estimating emission-line strengths.
A reduction in stare mode has also been carried out, where
the sky is estimated using a sigma-clipped mean in the regions free of host-galaxy light. 
The noise images for both reduction modes are constructed directly from
the data at each wavelength bin where the standard deviation is calculated in
the regions excluding the host. 

The positions of the SN images on the slit were determined by measuring the distance from the host nucleus in the {\it HST} image. H$\alpha$ emission from the host nucleus is clearly visible in the two-dimensional spectrum. 

\subsubsection{(a) Extraction and Combination}

We have extracted two separate sets of spectra to study the SN and to examine the strong nebular emission.
To study the SN light, we use a 6 pixel (1\farc08) wide aperture
centered at the positions of the SN.  We use a model of sky lines from
ESO\footnote{http://www.eso.org/observing/etc/skycalc/skycalc.htm} to construct a mask of strong night-sky emission.   In addition to line emission, the sky also produces continuum emission, and we model and subtract the continuum emission by computing the median value in bins of 200\,\AA\ in wavelength, after removing 
pixel elements with radiance $>$ 20,000 photons s$^{-1}$ m$^{-2}$ $\mu$m$^{-1}$ arcsec$^{-2}$.
After subtracting the continuum emission from the sky, we interpolate the
background-subtracted sky spectrum to the X-shooter wavelength grid and mask 
pixels with sky values greater than 10,000 photons s$^{-1}$ m$^{-2}$ $\mu$m$^{-1}$ arcsec$^{-2}$ when binning.
We also mask any pixels within 6.4\,\AA\ (5$\sigma$) of a strong nebular emission
line, including H$\alpha$ and \nii.
To study the host-galaxy narrow-line emission, we instead extract using an aperture with a width that is 250\% of the average FWHM seeing listed in Table~\ref{tab:seeing}. 

For both sets of extractions, the flux in each aperture is summed, and the noise spectrum within the same aperture is added in quadrature.
We next apply aperture corrections appropriate for the SN point source.
To estimate the expected slit loss for the SN and contamination from the early-type lens and host galaxy, we convolve a pre-explosion {\it HST} WFC3 $F160W$ image with a Gaussian kernel to produce 
an image having the average FWHM expected during each X-shooter observation (see Table~\ref{tab:seeing}). 
From the position angle and target positions, we next create pixel masks of the slit apertures. 
We use these masks and the convolved images to calculate the average expected SN
and galaxy light admitted through the slit. 
We correct each spectrum for the slit losses appropriate for the SN point source. 

We smooth each of the OBs separately and remove 5$\sigma$ outliers. 
The spectra are finally combined using a weighted average, and we propagate 
uncertainties.

\subsection{(4) Estimating the Luminosity of the SN H$\alpha$ Emission}
We measure the H$\alpha$ strength by fitting a Gaussian to the emission after subtracting 
a continuum level. We estimate a median flux in the wavelength range 14,500--15,500\,\AA\ for the grism spectrum. In the case of the X-shooter data, the continuum estimate is the median of the flux in the range 15,300--15,600\,\AA\ and 16,500--17,000\,\AA.  The uncertainty is computed by repeating this procedure to the bootstrap-resampled spectra. 

To assemble the comparison plot in Figure~\ref{fig:halphastrength}, we renormalize the SN~1987A-like SNe and SN 2005cp comparison spectra to match published total magnitudes in the papers listed in Table~\ref{tab:comparisondata}.
We next correct the spectra to remove Milky Way and host-galaxy reddening according to the values in Table~\ref{tab:comparisondata} and shift the comparison spectra to the rest frame.
The spectra of the low-redshift comparison SNe have high S/N, and
we estimate the continuum visually from the region adjacent to their P-Cygni
H$\alpha$ profiles while avoiding the \baii\ $\lambda$6142 feature.
 
\subsubsection{(a) Nebular Emission from Near the SN Position and Host Nuclear Region} 
We identify the host-galaxy emission lines \oii\ $\lambda$3727,
\hg, \hb, \oiii\ $\lambda$4959, \oiii\ $\lambda$5007, \ha, \nii\ $\lambda$6584,
\sii\ $\lambda$6716, and \sii\ $\lambda$6731 with varying statistical significance. 
Narrow lines are detected across much of the spatial direction of the slit,
which extends along both the host nucleus and position of SN images S1 and
S2 for observations OB1 and OB2 and the position of SN images S2 and S3
for OB3. As an illustration, Figure~\ref{fig:VLTexpsetup} shows the detection of \oii\ in OB2. 
The spatial dimension of the slit allows us to determine both the conditions local to the SN and within the nuclear region.

Parameters useful for the determination of the conditions of the emitting gas are the
fluxes and widths of the strong nebular emission lines. 
Strong-line diagnostics using the measured line fluxes can then be used to infer
the oxygen abundance of the emitting gas.

To estimate the uncertainty of the strong-line fluxes, we create bootstrapped spectra
from the set of four separate spectra of images of the explosion site assembled 
from the OB1 and OB2 observations of SN images S1 and S2. 
Unlike the spectra taken during OB3, the spectra acquired in OB1 and OB2 show
no evidence that emitting sources in the ``off regions'' were subtracted from the spectrum of the ``on regions.'' 

For each of the bootstrapped samples, the lines of \oii\ $\lambda$3727,
\hb, \oiii\ $\lambda$4959, \oiii\ $\lambda$5007, \ha, and
\nii\ $\lambda$6584 are fit using weighted least-squares, where for the \oii\ and \oiii\ doublets the widths are required to be the same, and the position and width of \nii\ are
determined from the fit to \oiii\ $\lambda$5007 which is not contaminated by night-sky emission. 
We use the inverse variance as the weight. 
To estimate the uncertainty of the individual fits,
we fit the spectrum assembled from each bootstrap combination 100 times after resampling within the uncertainty. We use the mean and the standard deviation of the resulting distribution to estimate the value and uncertainty of the line flux. 

We do not detect \nii\ in the spectrum extracted at the SN position, so we
instead report an upper limit. To determine the value of the
upper limit, we add an artificial line of increasing strength with a width matching that of the well-detected \oiii\ $\lambda$5007 and repeat until we obtain a 3$\sigma$ detection. 

For the host-galaxy nuclear region, we only have two extractions from OB1 and OB2, so we find a weighted combination of the two independent extractions. Because the strength of \oiii\ $\lambda$5007 is weak at the position of the host-galaxy nucleus, we use the width of \oii\ instead of \oiii\ as a model for \nii\ and let it vary within 3$\sigma$ of the best-fit values for \oii.  Using \oii\ or \oiii\ to model the \nii\ line profile has only a $\sim5$\% effect on 
measured emission-line ratios.

\begin{figure*}
\centering
\subfigure{\includegraphics[angle=0,width=7in]{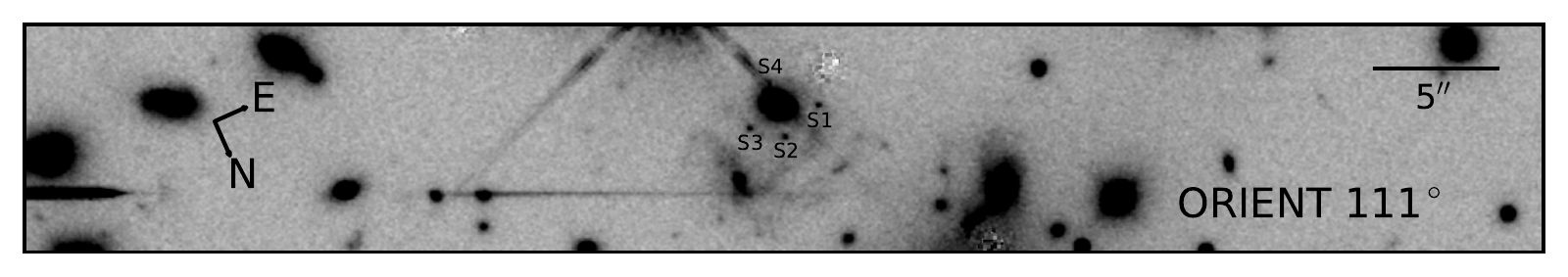}}
\subfigure{\includegraphics[angle=0,width=7in]{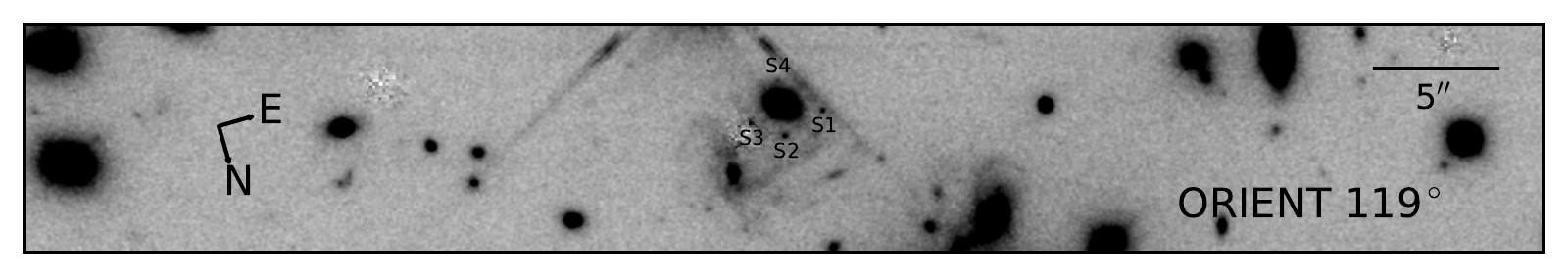}}
\subfigure{\includegraphics[angle=0,width=7in]{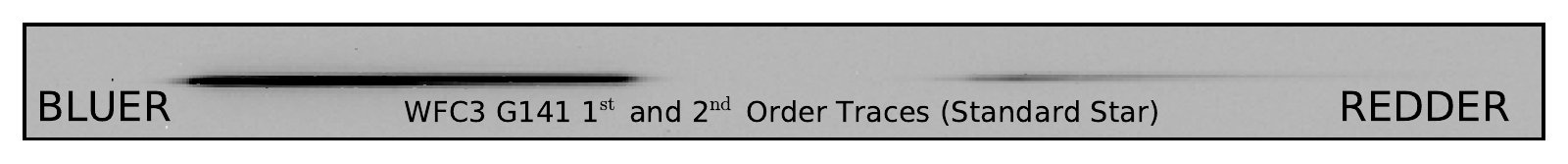}}
\caption{Coaddition of WFC3/IR $F125W$ and $F160W$ exposures taken
before G141 grism integrations at the same orientation. The dispersion axis is
horizontal. Unlike with a prism, blue light makes a smaller relative angle with
the grism. The first-order trace spans
11,000--17,000\,\AA\ ($\sim4400$--6800\,\AA\ in the SN rest frame). Bottom panel 
shows the first- and second-order traces in an observation of the standard star GD-153 (GO-13092; PI J.
Lee). }
\label{fig:grismorient}
\end{figure*}

\begin{figure*}
\centering
\subfigure{\includegraphics[angle=0,width=6.5in]{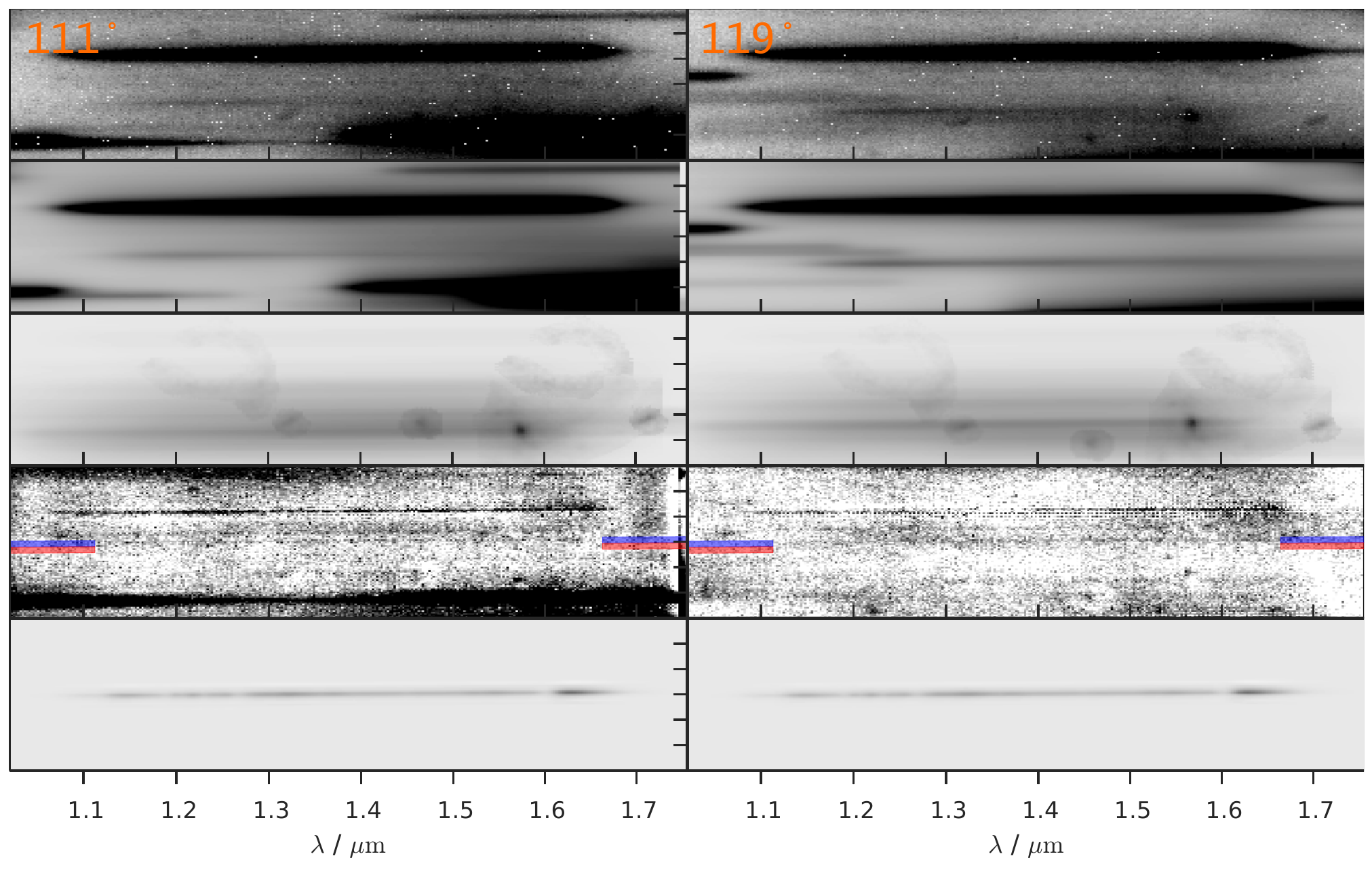}}
\caption{WFC3 G141 two-dimensional grism spectra taken of SN Refsdal in the
\one\ and \two\ orientations.  
For each orientation (beginning from the top), the first panel shows the full
observed spectrum, the second panel gives a model of nearby contaminants, the
third panel provides the model of the H$\alpha$ and \oiii\ emission
lines, and the fourth panel is the cleaned residual spectrum, with the stretch
now 10 times that of the preceding panels.  
The final panel is the model two-dimensional spectrum of the SN~IIP template.  
Blue bands show the extraction region of the S2 trace itself (it continues
across the whole spectrum but is only shown at the edges, so one can actually see
the spectrum under it) and the red band is the adjacent aperture used for the
local background.  The tick marks on the ordinate axis are spaced by 1\farcs
}
\label{fig:twod}
\end{figure*}

\begin{figure*}
\centering
\subfigure{\includegraphics[angle=0,width=6.5in]{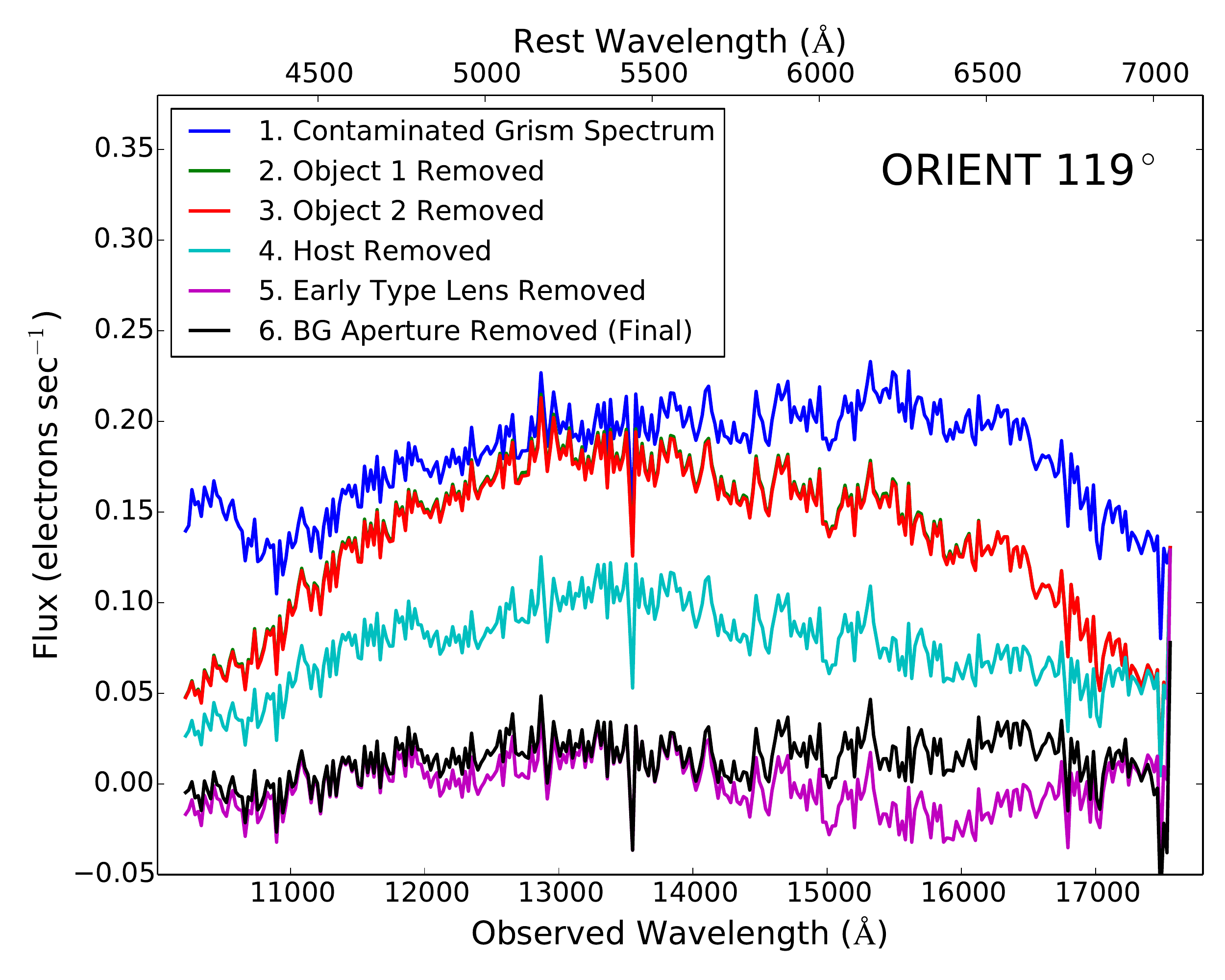}}
\caption{Successive subtraction of contamination along the WFC3 grism trace of
SN Refsdal taken in the
the \two\ telescope orientation.  The light from SN Refsdal constitutes only
10--15\% of the light 
in the extraction aperture. 
The steps follow in parallel the order of panels from top to bottom in 
Fig.~\ref{fig:twod}. The background subtraction (``BG Aperture'') is the flux
measured in an aperture 
running parallel to the extraction aperture. Background removal corrects a modest
overcorrection for
contamination at redder wavelengths. 
Despite extremely strong contamination, synthetic magnitudes of the subtracted
spectrum 
yield a $F125W-F160W$ color that has reasonable agreement with
magnitudes
measured from direct imaging (see Section~\ref{sec:specphot}). 
The level of contamination along of trace of image S2 in the grism spectrum
acquired at \one\ orient
is comparable.
}
\label{fig:contamination}
\end{figure*}

\begin{figure*}
\centering
\subfigure{\includegraphics[angle=0,width=6.25in]{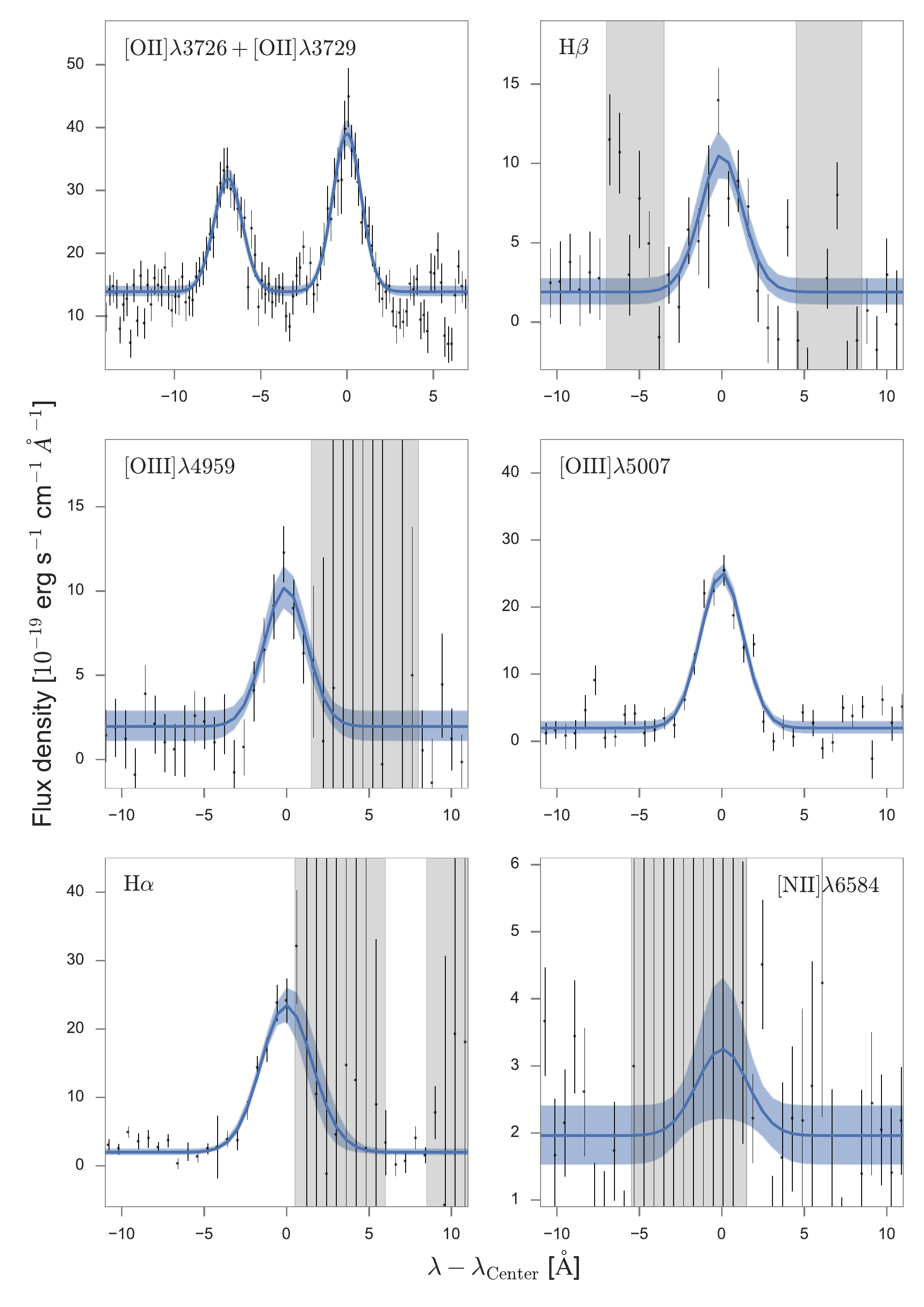}}
\caption{Profiles of narrow emission lines in X-shooter spectra at positions of 
SN images.   Each fitted model (blue) consist of a Gaussian
profile and a constant background.  Regions having strong sky background from emission lines are shown by the gray background. 
Here we first find $\mu$ and $\sigma$ that yield the best fit to the \oiii\ $\lambda$5007
emission line,
and, during fits to other
emission lines, we allow the line amplitude and local background to vary.  
The widths of the emission lines arising from forbidden transitions
(\oiii\ $\lambda\lambda$4959, 5007) and permitted transitions (H$\alpha$ and
H$\beta$) are comparable, which suggests that the narrow-line emission arises from 
\ionpat{H}{ii} regions instead of from the SN. 
}
\label{fig:emissionlines}
\end{figure*}

\end{document}